\documentclass[sigconf, natbib=false]{acmart}
\usepackage{amsmath}
\usepackage{cite}
\usepackage{verbatim}
\usepackage{graphicx}
\usepackage{subfig}
\usepackage{mdwlist}
\usepackage[noend]{algorithmic}
\usepackage{algorithm}
\usepackage{hyperref}
\usepackage{url}
\usepackage[T1]{fontenc}
\usepackage{epstopdf}
\usepackage{array}
\usepackage{arydshln}
\usepackage{mathtools}
\usepackage{balance}  
\usepackage{float}
\usepackage{enumitem}
\usepackage{xspace}
\usepackage{bbm}
\usepackage{multirow}
\usepackage{multicol}

\setlist{nolistsep,leftmargin=*}
\newtheorem{problem}{\textbf{Problem}}
\newtheorem{defn}{\textbf{Definition}}
\newtheorem{thm}{\textbf{Theorem}}
\newtheorem{corl}{\textbf{Corollary}}

\newtheorem{ex}{\textbf{Example}}

\newcommand{\ignore}[1]{}

\newcommand\imageWidthIcePlot{\textwidth*3/17}
\newcommand{\figcaption}[1]{\vspace*{-2.0mm}\caption{#1}\vspace*{-6mm}}
\newcommand{\tabcaption}[1]{\vspace*{-3mm}\caption{#1}\vspace*{-5mm}}

\copyrightyear{2019}
\acmYear{2019}
\setcopyright{iw3c2w3}
\acmConference[WWW '19]{Proceedings of the 2019 World Wide Web Conference}{May 13--17, 2019}{San Francisco, CA, USA}
\acmBooktitle{Proceedings of the 2019 World Wide Web Conference (WWW '19), May 13--17, 2019, San Francisco, CA, USA}
\acmPrice{}
\acmDOI{10.1145/3308558.3313448}
\acmISBN{978-1-4503-6674-8/19/05}

\title%[Relationship-Aware Graph Querying]
{RAQ: Relationship-Aware Graph Querying in Large Networks}

\author{Jithin Vachery}
\affiliation{%
  \institution{Computer Science and Engineering,}
  \institution{Indian Institute of Technology, Madras}
}
\email{jithin@cse.iitm.ac.in}

\author{Akhil Arora}
\affiliation{%
  \institution{EPFL}
}
\email{akhil.arora@epfl.ch}

\author{Sayan Ranu}
\affiliation{%
  \institution{Computer Science and Engineering,}
  \institution{Indian Institute of Technology, Delhi}
}
\email{sayanranu@cse.iitd.ac.in}

\author{Arnab Bhattacharya}
\affiliation{%
  \institution{Computer Science and Engineering,}
  \institution{Indian Institute of Technology, Kanpur}
}
\email{arnabb@cse.iitk.ac.in}

\begin{document}

\begin{abstract}
	The phenomenal growth of graph data from a wide variety of real-world
	applications has rendered graph querying to be a problem of paramount
	importance. Traditional techniques use structural as well as node
	similarities to find matches of a given query graph in a (large) target
	graph. However, almost all existing techniques have tacitly ignored the
	presence of \emph{relationships} in graphs, which are usually encoded
	through interactions between node and edge labels. In this paper, we
	propose \emph{RAQ---\textbf{R}elationship-\textbf{A}ware Graph
	\textbf{Q}uerying}---to mitigate this gap. Given a query graph, RAQ
	identifies the $k$ best matching subgraphs of the target graph that encode
	similar relationships as in the query graph. To assess the utility of RAQ
	as a graph querying paradigm for knowledge discovery and exploration tasks,
	we perform a \emph{user survey} on the Internet Movie Database (IMDb),
	where an overwhelming $86\%$ of the $170$ surveyed users preferred the
	relationship-aware match over traditional graph querying.  The need to
	perform subgraph isomorphism renders RAQ NP-hard.  The querying is made
	practical through \emph{beam stack search}.  Extensive experiments on
	multiple real-world graph datasets demonstrate RAQ to be effective,
	efficient, and scalable.
	%up to $5$ times faster than baseline strategies.
	%
\end{abstract}

\maketitle

\section{Introduction}
\label{sec:intro}

Graphs act as a natural choice to model data from several domains. Examples
include social networks~\cite{mywww}, knowledge graphs~\cite{freebase,dbpedia},
and protein-protein interaction networks (\emph{PPIs}) \cite{resling,reslingj}.
Consequently,
%The widespread usage of graph data has naturally necessisated the need to
%query graph data and RDF \cite{rdf} as the preferred choice for data
%interchange through the web,
graph-based searching and querying have received significant interest in both
academia \cite{NEMA, NESS, deltacon, saga,nbindex,edbt} and industry (e.g., Facebook's Graph
Search\cite{facebook} and Google's Knowledge Graph\cite{google}).
%require searching mechanisms that work with graph-based data rather than the
%traditional text-based queries. In such graphs, nodes denote entities and
%edges denote the relationships between entities. In most cases, the nodes are
%tagged with additional meta information to characterize the corresponding
%entities \cite{representative}. The ability to query these graph datasets
%efficiently is a fundamental necessity for a wide array of applications
%\cite{NEMA, NESS, representative, ctree}. 

\begin{figure}[t]
	\centering
	\subfloat[Query]{
	\label{fig:query}
	\includegraphics[width=3in]{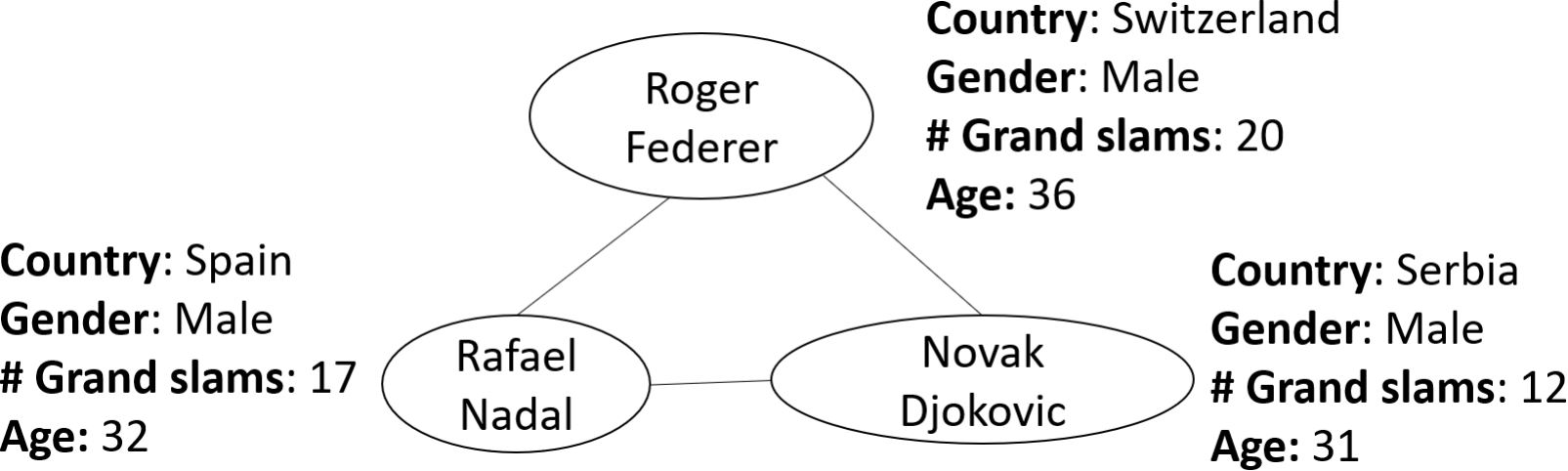}}\\
	\subfloat[Potential match]{
	\label{fig:raq}
	\includegraphics[width=3in]{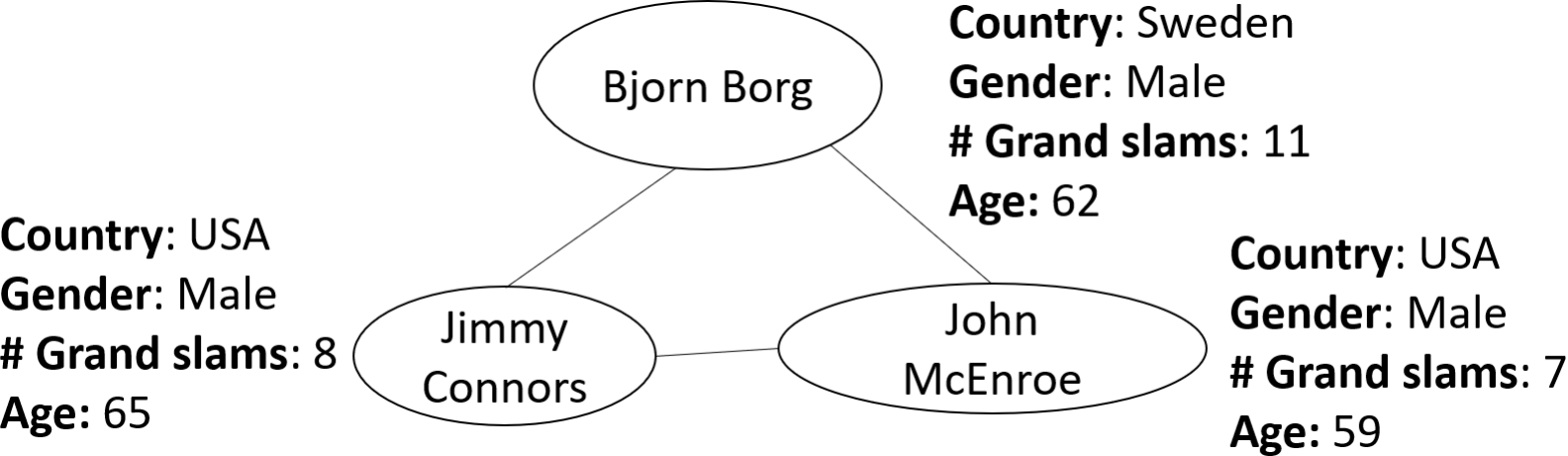}}
	\vspace{-0.05in}
	\figcaption{\textbf{Two subgraphs from the Wikipedia knowledge graph. For simplicity, we drop the directionality and model them as an undirected graph.}}
	\vspace{-0.1in}
\end{figure}

One of the most common queries in these frameworks is to find similar
embeddings of a query graph $q$ in a much larger target graph $G$. More
formally, given a query graph $q$, a target graph $G$ and a similarity function
$sim(q,g)$, the goal is to identify the top-$k$ most similar subgraphs $g
\subseteq G$ to the query graph $q$.  Traditional similarity functions consider
two graphs as similar if they are structurally similar and they contain similar
nodes\cite{NEMA, NESS, ged_similarity1,node_embedding_similarity}. Two nodes
are said to be similar if they are represented by similar feature vectors; the
most common form being just a node label. Despite significant advancements in
the field of graph querying\cite{deltacon,NEMA,ged_similarity1}, this commonly
adopted definition of similarity is oblivious to the presence of
\emph{relationships} in the graphs.  In this work, we bring the power of
relationships to graph querying.

To give an example of a relationship, consider the entity tuples
$\langle$Rafael Nadal, Roger Federer$\rangle$ and $\langle$John McEnroe, Bjorn
Borg$\rangle$ from the Wikipedia knowledge graph. Although the entities in
these two tuples are different, both tuples encode a similar relationship; they
represent players who played during the same era and won multiple grand slam
tournaments between them. Annotating knowledge graphs with relationship
information in the form of edge labels has been shown to expose a higher level
of semantics than simple entity-to-entity matching\cite{exemplar2,acl}.
%extend this feature to

%To illustrate the importance of relationships, consider the Wikipedia
%knowledge graph, where each node corresponds to a Wikipedia page denoting an
%entity, and there is an edge from node $u$ to $v$, if $u$'s webpage contains a
%hyperlink to $v$'s page. 
Fig.~\ref{fig:query} presents a query subgraph from Wikipedia knowledge graph
containing well-known tennis players of the current era. Wiki\-pedia further
characterizes each node with a bunch of summary features as shown in the
figure.  Now, consider the task of identifying a subgraph from Wikipedia
similar to the query in Fig.~\ref{fig:query}. Consider Fig.~\ref{fig:raq},
which depicts another group of well-known tennis players: Bjorn Borg, Jimmy
Connors and John McEnroe. Would a tennis fan consider these two groups to be
similar?  There is a strong possibility that she would since both the groups
respectively represent players who played during the same era and won multiple
grand slam tournaments among them. 
%In Fig.~\ref{fig:raq2}, we present yet another subgraph of Wikipedia
%containing female tennis players, who are connected by the same relationships.

%The above observation leads us to an important question: \emph{Would existing
%graph similarity functions capture this human thought process and consider the
%shown subgraphs as similar?} 
Let $g_1$ and $g_2$ be two graphs. Most of the existing techniques \cite{NEMA,
ctree, ged_similarity1, ged_similarity2, node_embedding_similarity} rely on
mapping each node in $g_1$ to some node in $g_2$, and the cumulative similarity
of the mapped node pairs dictate the overall similarity score between $g_1$ and
$g_2$. When such a strategy is applied to the above two subgraphs
(Fig.~\ref{fig:query} and~\ref{fig:raq}), none of the possible node mappings
would return a high score since their feature vectors are not similar. For
example, Federer is not similar to any of the nodes in Fig.~\ref{fig:raq},
since he is from a different generation as indicated by his ``Age'', and
belongs to a different country. The ``Gender'' and ``\# Grand slams'' will
fetch some similarity, however, that may not be enough to offset the
dissimilarity in the other two features. More fundamentally, although a graph
contains a \emph{group/collection} of nodes, \emph{existing techniques} treat
each node as an \emph{independent entity} while comparing two graphs.
Consequently, \emph{group-level relationships}, that can be identified only by
analyzing the \emph{group-level interactions} between node feature vectors, are
not reflected in the similarity score. For example, the relationship that each
group contains players from roughly the same generation, as hinted by the
``Age'' feature, cannot be inferred through
node similarity alone, where nodes are treated independently.  This observation
forms the core motivation behind our work to move beyond node similarity, and
match subgraphs based on the \emph{relationships encoded through non-trivial
interactions between node features}.

Several important questions arise at this juncture. \textit{What is the precise
definition of a relationship? How do we quantify the strength of a
relationship? Are all relationships equally important?} In this paper, we
answer these questions. In sum, we make the following key contributions:

\begin{itemize}[nosep,leftmargin=*]	

	\item We propose a novel graph querying paradigm, \emph{RAQ
		(Relationship-Aware Graph Querying)}, that incorporates the notion of
		\emph{relationship} in its similarity function
		(Sec.~\ref{sec:problem}). The proposed formulation automatically mines
		the relationships manifested in the query graph and ranks their
		importance based on their \emph{statistical significance}
		(Sec.~\ref{sec:relationship}).

	\item We design the RAQ search algorithm, which employs \emph{beam stack
		search} on an R-tree to identify good \emph{candidates}. These
		candidates are further refined through \emph{neighborhood signatures}
		constructed using random walks with restarts to prune the search space
		and compute the \emph{optimal} answer set. (Sec.~\ref{sec:algo}).

	\item Empirical evaluation (Sec.~\ref{sec:exp}) on real datasets
		establishes three key properties. (1) RAQ produces results that are
		\emph{useful}.  This is validated by two user surveys.
		%spanning $170$ users conducted on the Internet Movie Database (IMDb),
		%where $81\%$ users deemed the ranking of relationship importance using
		%statistical significance to be correct, and an overwhelming majority
		%of $86\%$ users preferred the weighted relationship-aware similarity
		%function.
		(2) RAQ \emph{complements} existing graph similarity metrics by finding
		results that traditional techniques are unable to find. (3) RAQ is up
		to $5$ times faster than baseline strategies and scales to
		million-sized networks.
		
\end{itemize}

The code is available at:
\url{https://github.com/idea-iitd/RAQ}.

\begin{comment}
\begin{table}[b]
\centering
\small
\scalebox{0.8}{
	\begin{tabular}{|c|c|}\hline
		\textbf{Item} & \textbf{Definition}\\\hline
		\hline
	%	$V$ & Set of target graph vertices; $|V| = n$. \\\hline
	%	$E$ & Edge set for the nodes in $V$; $|E| = m$.\\\hline
	%	$V_q$ & Set of query graph vertices; $|V_q| = n_q$. \\\hline
	%	$E_q$ & Edge set for the nodes in $V_q$; $|E_q|= m_q$.\\\hline
	%	$F$ & Set of node feature types; $|F|=d$.\\\hline
	%	$G=(V,E,\mathbb{F})$ & Labeled target graph.\\\hline
	%	$q=(V_q,E_q,\mathbb{F}_q)$ & Labeled query graph.\\\hline
        $g_1 \subseteq g_2$ & $g_1$ is subgraph isomorphic to $g_2$\\\hline
		$f(v)=[f_1(v),\ldots,f_d(v)]$ & $d$-dimensional feature vector for node $v$\\\hline
		$\mathcal{N}(f_i,e)$ & Neighborhood vector for feature $f_i$ around edge $e$\\\hline
		$w(e)=[w(f_1,e),\ldots,w(f_d,e)]$ & Weight vector of edge $e$\\\hline
		$s(e)=[s(f_1,e),\ldots,s(f_d,e)]$ & Relationship vector corresponding to edge $e$\\\hline
		$eSim(e,e')$ & Contextual similarity between edges $e$ and $e'$\\\hline
		$\phi: V_q \rightarrow V$ & (Sub)graph injection function\\\hline
		$RAQ_{\phi}(q,G)$ & Contextual similarity between $q$ and $G$\\\hline
	\end{tabular}
}
\tabcaption{\textbf{Summary of the notations used.}}
\label{tab:terminology}
\moveups
\end{table}
\end{comment}
\section{Related Work}
\label{sec:qbe}
%The general problem of graph querying has been studied extensively over the past decade. 
%Thus, it is rather difficult to write a complete literature review in one section. 
%Here, we overview the existing works that overlap with our problem. We skip the comparison with Query-by-example paradigm since it has already been done in Sec.~\ref{sec:qbe}.
Research work done in both exact and approximate (sub-)graph querying have employed a plethora of similarity functions. The most prominent of them being graph edit distance (GED) \cite{ged_similarity1,ged_similarity2}, maximum and minimum common subgraph \cite{mcg_similarity1,mcg_similarity2}, edge misses \cite{tale}, structural similarity \cite{best_effort,saga,isorank,NESS,NEMA,deltacon}, node-label mismatches \cite{saga,isorank}, and statistical significance of node-label distributions \cite{graphquerying_chi}. However, all these methods operate oblivious to the presence of \emph{relationships in the query graph}.

\textbf{Query-by-example Paradigm: }
The need to query based on relationships is touched upon by the recent line of work on Query-by-example\cite{exemplar,exemplar1,exemplar2,exemplar3}. All these techniques take an edge or a set of edges as input, and this input is treated as an \emph{example} of what the user wants to find. Each exemplar edge connects two entities of the knowledge graph and the edge label denotes the relationship between these entities. Typical relationships in a knowledge graph are ``friendOf'', ``locatedIn'' etc. Given the exemplar edges, the common goal in this line of work is to identify subgraphs from the target graph that are connected by the same edge labels as expressed in the exemplars. Our work is motivated by the same line of thought and revamps this paradigm through the following innovations.
\begin{enumerate}
\item \textbf{Mining relationships: } While \cite{exemplar,exemplar1,exemplar2,exemplar3} assume that relationships are explicitly provided in the form of edge labels, we \emph{mine} these relationships from feature-value interactions observed between a pair of nodes. Thus, RAQ is not constrained by the explicit availability of edge relationships. This results in a more powerful model since it is hard to capture all possible relationships through a small set of edge labels. Furthermore, unlike existing query-by-example techniques, RAQ is not limited to edge-labeled (relationship-annotated) graphs alone, and exposes a more generic framework capable of operating on graphs with both node and edge labels.
\item \textbf{Multiple relationships: }Existing techniques assume each edge to be associated with only one relationship. In our formulation, each edge encodes $d$ relationships corresponding to each of the $d$ features characterizing a node. 
\item \textbf{Relationship importance: } All relationships in the exemplar edges are assumed to be of equal importance. In the proposed paradigm, we analyze all of the identified relationships and quantify their importance through statistical significance. As we will see later in Sec.~\ref{sec:survey}, users find this weighting relevant.
\item \textbf{Relationship similarity: }Existing query-by-example techniques are incapable of computing similarity between relationships. This is a natural consequence of the fact that relationships are reduced to \emph{categorical edge labels}. Thus, relationships between two pairs of nodes are either considered \emph{identical or different}.\ignore{We model relationships as numbers and consequently} In contrast, \emph{RAQ is not limited to a binary matching}. For example, the relationship with respect to ``\# Grand slams'' between Federer and Nadal is similar to that between Bjorg and McEnroe, but not identical.
\end{enumerate}

Overall, the proposed paradigm is more generic, and therefore more challenging to solve algorithmically. %This is substantiated by results presented in Sec.~\ref{sec:survey}, where the RAQ search algorithm achieves significant improvement in empirical performance over traditional node similarity and query-by-example techniques.

\section{Problem Formulation}
\label{sec:problem}

%In this section, we formulate the problem of relationship-aware graph querying and
%define the key concepts.  
%The basic objective of our problem is to identify similar
%relationship-aware embeddings of a query graph $q$ in a (large) target graph $G$. A graph is \emph{relationship-aware} similar to $q$ if it preserves the same relationships and context expressed in $q$. The notations used in this paper are summarized in Table~\ref{tab:terminology}.

%\subsection{Preliminaries}
The input to our problem is a \emph{target graph}, a \emph{query graph}, and the value $k$ corresponding to the top-$k$ matches that need to be identified.
\begin{defn}[Target Graph]
	\label{def:target_graph}
	The target graph $G = (V,E,\mathbb{F})$ contains a node set
	$V$, an edge set $E$, and a set of $d$-dimensional feature vectors $\mathbb{F}$, where each feature vector $f(v)=[f_1(v),\ldots,f_d(v)]$ characterizes a node $v\in V$. Optionally, each edge may also be annotated with an edge label.
%, where (1)
%	each target node $v \in V$ represents an entity, (2) each target edge
%	$e=(u,v) \in E$ denotes the relationship between the two entities $u$ and
%	$v$, and (3) 
%$\mathbb{F}$ is a set of $d$-dimensional feature vectors
%	$f(v)=[f_1(v),\ldots,f_d(v)]$ associated with each node $v \in V$.
	%
\end{defn}

Without loss of generality, we assume that for each feature value $f_i(v)$, $0 \leq f_i(v) \leq 1$.
A query graph $q = (V_q,$ $E_q$, $\mathbb{F}_q)$ follows the same notational structure as the target graph. % where nodes $V_q$ and $E_q$ represent the set of nodes and edges in the query graph, and $\mathbb{F}_q$ characterizes the query nodes. 
Both the target graph and the query graph can be either directed or undirected. 
%In our framework, the user or application only provides the structure of the query
%graph, i.e., the nodes and edges, and does not require any knowledge of the
%feature vectors. The feature vectors of the nodes are extracted automatically
%from the data repository. For example, DBLP could serve as the data repository
%in co-authorship networks.
%In practice, the target graph would normally represent a large network such as a knowledge graph\cite{freebase, dbpedia}, collaboration network\cite{dblp}, or a protein-protein interaction network\cite{ppi}. The feature vectors represent node attributes as discussed in Sec.~\ref{sec:intro}. The query graph would normally be a much smaller graph, where the goal is to extract subgraphs of the target graph that are \emph{similar} to the query graph. Therefore, we are looking for structurally 
Our goal is to identify \emph{isomorphic} embeddings of the query graph within the target graph, such that they encode relationships similar to that manifested in the query. 
%\begin{defn}[Graph Isomorphism]
%\label{def:isomorphism}
Graph $g_1=(V_1,$ $E_1,$ $\mathbb{F}_1)$ is \emph{isomorphic} to
$g_2=(V_2,E_2,\mathbb{F}_2)$ if there exists a bijection $\phi$ such that
for every vertex $v \in V_1,\; \phi(v) \in V_2$ and for every edge $e =
(u, v) \in E_1, \phi(e)=(\phi(u), \phi(v)) \in  E_2$ and the labels of $e$ and $\phi(e)$ are the same.
%
%\end{defn}
%Note that there may exist multiple isomorphic mappings between two graphs. One such example are the graphs in Fig.~\ref{fig:query} and Fig.~\ref{fig:raq}.  
Among all isomorphic embeddings of the query graph $q$ in the target graph $G$, we want to select the $k$ embeddings that possess the highest \emph{relationship-aware similarity (RAS)} to $q$ (defined formally in Def.~\ref{def:ras}). 
%The concept of \emph{subgraph isomorphism} is defined analogously by using an
%\emph{injection} instead of a bijection. We use the notation $g_1 \subseteq
%g_2$ to denote that $g_1$ is subgraph isomorphic to $g_2$.
\begin{problem}[Top-$k$ Relationship-aware Querying]
	Given a query graph $q$, a target graph	$G$, and a positive integer $k$, identify the graphs
	$\mathbb{A}=\{g_1, \cdots, g_k\}$, such that
%	\begin{enumerate}
	    (1) $\forall g_i\in \mathbb{A}, g_i$ is a subgraph of  $G$, (2) $\forall g_i\in \mathbb{A}, g_i$ is isomorphic to $q$, and (3) $\nexists g'\subseteq G$ such that  $RAS(g',q)>RAS(g_i,q)$ for some $g_i\in\mathbb{A}$. 
	%\end{enumerate}
\end{problem}

The main task is to design the relationship-aware similarity function
$RAS(g,q)$ for any two isomorphic graphs $g$ and $q$.

%While we mostly discuss top-$k$ queries, our similarity function as well as the
%proposed index structure can easily be applied to range queries. A discussion
%on range queries is included in Appendix~\ref{sec:rangequery}. We focus on top-$k$
%query since specifying $k$ is more intuitive for the end user than a similarity
%threshold. 
\section{Relationship-aware Similarity}
\label{sec:relationship}

We first define what a relationship is. 
Intuitively, each feature represents the characteristics of a node and the
relationship between two nodes is inferred by studying the interplay between
their feature values. For example, considering
Djokovic and Nadal, both entities are of the same gender, similar age, have won
a large number of grand slam tournaments, but are citizens of different countries.
%It must be noted that a relationship should not be equated to only homogeneity
%in feature values. Diversity in feature values may itself be a desired
%relationship. For example, in a collaboration network, one may want to match
%groups based on the multi-disciplinary nature of their expertise areas. In
%other words, each feature characterizes a node, and the interplay between the
%feature values of nodes that are connected by an edge forms a relationship. 
%Each edge encodes the relationships between the two nodes it connects. 
Interactions between each corresponding feature of the connected nodes
generate a relationship, and hence, a $d$-dimensional feature vector would
produce $d$ relationships per edge. 
%To give an example, the edge between Djokovic and Nadal in
%Fig.~\ref{fig:query} encodes the relationships that both entities are of same
%gender, similar age, have won a large number of grand slam tournaments, and
%citizens of differing countries. 
We capture these relationships in the form of a \emph{relationship
vector} $s(e)$ corresponding to each edge $e=(u,v)$.

\begin{defn}[Relationship Vector]
%\begin{defn}[Relationship Vector]
  \label{def:rv}
	\textit{The relationship vector, $s(e)$, of an edge $e = (u,v)$ captures the features
	that are preserved in the relationship between $u$ and
    $v$. Formally, $s(e) = [s(f_1,e),\ldots,s(f_d,e)]$ where }
	\begin{align}
		\nonumber
		%\nonumber
		%
	    s(f_i,e) &=
			\begin{cases}
				\Gamma(f_i(u),f_i(v)), \quad \text{if} \quad f_i \text{ is real valued} \\
				1, \quad \text{if} \quad f_i \text{ is categorical and } f_i(u) = f_i(v) \\
				0, \quad \text{if} \quad f_i \text{ is categorical and } f_i(u) \neq f_i(v)
		    \end{cases}
	\end{align}
	\begin{align}
		\label{eq:J}
		%\text{where } \Gamma(x,y) &= \frac{\min\{x,y\}}{\max\{x,y\}}
		\text{where } \Gamma(x,y) &=
			\begin{cases}
				1, \quad \text{if } x=y=0\\
				\min\{x,y\} / \max\{x,y\},\quad \text{otherwise }
			\end{cases}
	\end{align}
\end{defn}

\begin{ex}
	\label{ex:relationship}
	\textit{The relationship vector $s(e)$ for the edge $e$ connecting Nadal and Djokovic in Fig.~\ref{fig:query} is $[0, 1, 0.71, 0.97]$. The second dimension is $1$ since it is a categorical feature and both are males. The third dimension is non-categorical and thus, the similarity between them is $\frac{12}{17}$. Similarly, the relationship vector of the edge between Connors and McEnroe in Fig.~\ref{fig:raq} is $[1,1,0.88,0.91]$.}
\end{ex}

From its definition, each dimension in the relationship vector takes a value in the range $[0,1]$. A value close to $1$ indicates that both endpoints of the edge share similar values. Two edges $e$ and $e'$ are considered similar if they encode similar relationships. Thus, to quantify their similarity, we compute the similarity between their relationship vectors:
\vspace{-0.10in}
\begin{align}
	\label{eq:edgesim}
	eSim(e,e')& = \sum_{i=1}^{d} w(i) \cdot \Gamma\left( s(f_i,e),s(f_i,e')\right)
\end{align}

$eSim(e,e')$ is a weighted \emph{min-max} similarity between the two
relationship vectors. The \emph{weight} $w(i)$ of the $i^\text{th}$ relationship
(feature) represents its relative importance with respect to the other
relationships, and the function $\Gamma\left( s(f_i,e),s(f_i,e')\right)$
operates as defined in Eq.~\ref{eq:J}. We add the constraint that $\sum_{i=1}^d
w_i=1$.
\begin{ex}
	\label{example:edge_matching}
	Assume that each relationship is of equal importance and, therefore, $\forall i$, $w_i=\frac{1}{d}$. The similarity between the edges Nadal-Djokovic and Connors-McEnroe is, thus, $0.25 \times 0 + 0.25 \times 1 + 0.25 \times	\frac{0.71}{0.88} + 0.25 \times \frac{0.91}{0.97} = 0.68$.
\end{ex}

Given two isomorphic subgraphs $q$ and $g$, and an isomorphic mapping $\phi$
from the edges in $q$ to the edges in $g$, the \emph{relationship-aware similarity (RAS)}
with respect to the mapping $\phi$ is the similarity between the mapped edges
from $q$ to $g$:
\begin{equation}
		RAS_{\phi}(q,g) = \sum_{\forall e_q=(u_q,v_q)\in E_q} eSim\left (e_q, \phi(e_q)\right)
\end{equation}
%between two edges, we are now ready to define relationship-aware similarity between two graphs.

\begin{defn}[Relationship-aware Similarity]
	\label{def:ras}
	The \\relationship-aware similarity (RAS) between $q$ and $g$ is the maximum RAS under all possible isomorphic mappings:
	\begin{align}
		\label{eq:graph}
		RAS(q,g) = \max_{ \phi}\left\{RAS_{\phi}(q,g) \right\}
	\end{align}
\end{defn}

%\subsection{Quantifying Relationship Importance}
\noindent\textbf{Quantifying Relationship Importance:}
The obvious question now is how to infer
the weights in $eSim(e,e')$ (Eq~\ref{eq:edgesim}). One option is to ask the
user to provide the weights. However, such a strategy may not be practical.
Specifically, each edge may encode a large number of relationships
corresponding to features characterizing the participating nodes, and providing
a weight with respect to each relationship is hard for
general users. Also, providing a large number of input parameters
is a cumbersome procedure. Hence, ideally, the weights should be automatically
inferred. 

The simplest approach is to assume that all relationships are of
equal importance, i.e. $\forall i, w_i=\frac{1}{d}$. This assumption however,
may be unrealistic. To elaborate, \emph{is the relationship that both Nadal and
Djokovic are males as important as the relationship that both of them have won
a large number of grand slam tournaments?} Intuitively, such an assumption
appears to be odd.

More importantly, should the importance be constant under all
situations and should it ignore the feature values. For example, consider an edge between
two tennis players who have won no grand slams. In such a case, both would have
a similarity of $1$ in the ``\# Grand Slams'' dimension. In that case, is the
importance of grand slams the same between these two players as compared to
Nadal and Djokovic? From a human psychological point of view, we value events
that are rare. The relationship between Nadal and Djokovic stands out since
they have won so many grand slams. We capture this intuition through
\emph{statistical significance}. Higher the statistical
significance of a relationship, more is its weight.

%As we will discuss later in
%Sec.~\ref{sec:survey}, our empirical evaluation through a user survey spanning
%$170$ users reveals an overwhelming preference towards the significance
%weighted model when compared to the assumption of uniform weights.

%Ex.~\ref{example:edge_matching} describes the relationship-aware similarity computation between two edges using the co-authorship network use-case as mentioned in Section~\ref{sec:intro}.

%Intuitively, the relationship vectors for both edges in Ex.~\ref{ex:relationship} are similar. However, as discussed in Sec.~\ref{sec:intro}, all relationships may not be equally important. Our next goal is therefore to learn the importance of each relationship based on the context prevalent in the query. 

%Note the node similarities do not play a role in our setting, and the focus is on
%identifying edges that are similar

\subsection{Statistical Significance of Relationships}
\label{sec:context}

Statistical significance tests quantify whether an event occurred due to chance
alone or is a result of some additional factor. Several statistical tests
exist \cite{stattests}, such as the chi-square test\cite{graphchi}, p-value\cite{graphsig,pgraphsig,graphsig_jcim},  g-test\cite{leap}, Kruskal-Wallis test, Anderson-Darling test, Shapiro-Wilk test, etc., to quantify the statistical
significance of an event. However, none of the techniques work directly on both
categorical data (e.g., country) as well as continuous-valued data (e.g., age).
Thus, we need to either map continuous data into a discrete space or convert
discrete, categorical data into a continuous domain. Since converting
categorical data such as gender and country into a continuous space is not
feasible, we discretize continuous data into bins. We use the \emph{chi-squared
statistic} \cite{chi_square} to measure statistical significance
since it is known to produce robust results
\cite{graphchi,stringchi} and is efficient to compute. 

Converting continuous data into discrete bins has been extensively studied
\cite{survey}. While any discretization
technique can be used, we use kernel-density estimation based method
\cite{binning} since it is unsupervised and
non-parametric and, therefore, easy to apply on any continuous valued variable.

Let $x$ be some event in a random experiment and $S$ be the sample space. For
all outcomes $y \in S$, let $O_y$ represent the number of times outcome $y$ has
occurred. The chi-square statistic, $X^2$, measures the statistical significance of the
event $x$ through the deviation of the \emph{observed} frequency of all
possible outcomes $O_y$ from their \emph{expected} frequency $E_y$ given by the
\emph{null model}:
\begin{align}
X^2 = \sum_{\forall y \in S} \frac{(O_y - E_y)^2}{E_y}.
\end{align}
The higher the chi-square, the more statistically significant the observed event is. 

In our problem, an event corresponds to a relationship. A relationship is manifested in the form of feature values assumed by the two endpoints of an edge. Thus, we should compute the observed frequency of a relationship and compare it to its expected frequency as per the chi-square protocol. Towards that end, each edge $e=(u,v)$ gives rise to a \emph{relationship tuple} $\tau=(f_i(u),f_i(v))$ corresponding to feature $f_i$. Let $\mathcal{V}(f_i)$ represent the set of unique tuples assumed by feature $f_i$ across all edges in a graph $g$. Then, the \emph{observed frequency}, $O_{\tau}$, for any relationship tuple $\tau \in \mathcal{V}(f_i)$ is simply the number of times $\tau$ is encountered in $g$.

\begin{ex}
Let us revisit Fig.~\ref{fig:query}. As a pre-processing step, we would first discretize ``\# Grand slams'' and ``Age'' into discrete categories. Let us assume that grand slam wins are bucketed into the following bins -- Poor: 0, Good: 1-5, Very Good: 6-10, Great: 11-15, All time great (ATG):$>15$. Therefore, $\mathcal{V}$( ``\# Grand slam'') contains the tuples (ATG,ATG) and  (ATG,Great) with observed frequencies 1 and 2 respectively. Similarly, $\mathcal{V}$(``Gender'') $=\{(Male, Male)\}$, where observed frequency $O_{(Male, Male)}=3$.
% and their obersver frequenciesfeature ``Gender'' takes only one value ``Male'', and therefore the observed frequency of ``Gender: Male'' is 3. On the other hand, ``Country'', ``Age'' and ``\#Grand slams'' contains three unique values, with each value possessing an observed frequency of $1$. 
\end{ex}

\noindent\textbf{Null Model:}
%Next, we formalize the null model so that the expected frequency of a tuple can be computed. 
%Towards that, we build the null model from the distribution of relationship tuples in the target graph. 
We assume that relationship tuples in the query graph are drawn independently and randomly from the distribution of tuples in the target graph $G$.
%, which would typically be a large network such as Wikipedia knowledge base, IMDb, etc. 
If the distribution of a relationship tuple in the query graph deviates significantly from the distribution in target graph, we call the corresponding relationship statistically significant. 
%To give an example, if the $(ATG,ATG)$ tuple in the target graph is extremely rare, and yet we find a high proportion of $(ATG,ATG)$ tuples in the query graph, then the ``\# Grand slam'' relationship is statistically significant. 

Let $P(f_i) =\{p_1, \ldots,$ $p_{|\mathcal{V}(f_i)|}\}$ denote the probability distribution of relationship tuples for feature $f_i$ in the target graph $G=(V,E,\mathbb{F})$. Here, $p_j\in P(f_i)$ represents the probability of finding tuple $\tau_j\in \mathcal{V}(f_i)$ in the target graph and is equal to the ratio $\frac{freq(\tau_j)}{|E|}$, where $freq(\tau_j)$ denotes the frequency of tuple $\tau_j$ in the target graph.  %from its value set $\mathcal{V}(f_i)$, i.e., $p(f_i(v)=\tau_j) \mid v \in V$, $\tau_j \in \mathcal{V}(f_i)$.
%For a query graph $q$ and a feature $f_i$, let us consider the unique feature
%values $\mathcal{V}(f_i,e)$, and the observed neighborhood vector
%$\mathcal{N}(f_i,e)$, where $\sum_{\tau \in \mathcal{V}(f_i,e)} \mathcal{P}(\tau)
%= \hat{n}$. 
Since each relationship tuple of feature $f_i$ in the query graph is drawn randomly and independently from $P(f_i)$, the \emph{expected frequency} $E_{\tau}$ of any tuple $\tau$ in the query graph $q=(V_q,E_q,\mathbb{F}_q)$ is
\begin{equation}
E_{\tau} = |E_q|. p_{\tau}%, \forall \tau \in \mathcal{V}(f_i,e)$ where
\end{equation}
The chi-square value of $f_i$ with observed frequency $O_{\tau}$ is, therefore,
\begin{align}
	\label{eqbasic}
	X^2_{f_i} = \sum_{\forall \tau \in \mathcal{V}(f_i)} \frac{(O_{\tau} - E_{\tau})^2}{E_{\tau}}
\end{align}
\begin{ex}
	\label{example:context_learn}
	Let us compute the significance of the grand slam and gender relationships
	in Fig.~\ref{fig:query}. Since very few of the tennis players have won a
	grand slam tournament, we assume that in the target graph, $5\%$ of the
	edges have the relationship tuple (ATG, Great) and $1\%$ have the tuple
	(ATG, ATG). Therefore, the expected frequency of (ATG, Great), (ATG, ATG)
	and all other tuples in Fig.~\ref{fig:query} is $0.05\times 3 = 0.15$, $0.01\times
	3 = 0.03$ and $0.94\times 3 = 2.82$ respectively.
	Consequently, 
\begin{equation}
\nonumber
X^2_{\# Grand slam}=\frac{(2-0.15)^2}{0.15}+\frac{(1-0.03)^2}{0.03}+\frac{(0-2.82)^2}{2.82}=57
\end{equation}

In the gender relationship, let us assume that in the target graph $50\%$ of the
edges contain the tuple (Male, Male), $40\%$ contain (Female, Female) and
$10\%$ contain (Female, Male). Therefore, the expected frequency of
(Male, Male), (Female, Female) and (Female, Male) in Fig.~\ref{fig:query}
are
$1.5, 1.2$ and $0.3$ respectively. Consequently, $X^2_{Gender}=3$. Therefore, we
conclude that the grand slam relationship is much more significant than the
relationship that all players are males. %This statistical inference conforms with the human intuition.
\end{ex}
Since the importance of each relationship is proportional to its statistical significance, the weight $w_i$ of the $i^\text{th}$ feature in Eq.~\eqref{eq:edgesim} is
\begin{align}
	\label{eq:contextweight_chi}
	w_i = \frac{X^{2}_{f_i}}{\sum_{j=1}^d X^{2}_{f_j}}
\end{align}

%The formalization of the relationship-aware similarity function (Eq.~\ref{eq:graph}) is now complete. %We therefore proceed to the next step of top-$k$ querying using the proposed similarity function.

%\section{Top-$k$ Relationship-aware Querying (RAQ)}
\section{Relationship-Aware Querying (RAQ)}
\label{sec:algo}

The na\"ive approach to solve a top-$k$ query is to first enumerate all possible
subgraphs of the target graph, identify those that are isomorphic to the query,
and then rank these isomorphic subgraphs based on their relationship-aware
similarity. However, this approach is prohibitively expensive since the target
graph contains exponential number of subgraphs. In fact, the problem is
NP-complete since finding isomorphic embeddings of a query graph within the
target graph reduces to \emph{subgraph isomorphism}\cite{ctree}. We, therefore,
need to design a more scalable searching algorithm.
% that is efficient and scalable to large graphs containing millions of nodes
% and edges.

\subsection{Bottom-Up Exploration}
We first analyze a bottom-up exploration strategy. First, we pick an
arbitrary edge $e_q$ from the query graph $q$ and map it to an arbitrary
edge $e$ of the target graph. We call $e_q$ the \emph{seed edge}. This forms an $1$-edge common subgraph of $q$ and
$G$. Next, we try to grow it into a $2$-edge subgraph by extending both $e_q$ and
$e$ through one of their neighboring edges such that the two extended subgraphs
are isomorphic and, therefore, a common subgraph of $q$ and $G$. This procedure
of growing using a neighboring edge continues iteratively till either of the following two cases become true:
\begin{enumerate}
\item The subgraph cannot be grown any further such that it remains a subgraph of both $q$ and $G$. In this case, we fail to find an isomorphic embedding of $q$.
\item The subgraph becomes isomorphic to $q$. We compute its similarity to $q$ and store it as a candidate.
\end{enumerate}

Once the growing procedure terminates, the search
restarts by mapping the seed query edge to another edge of $G$.

Once all
possible mappings of the seed edge are explored, we pick another query edge as
the seed and repeat the procedure. Finally, the $k$ highest scoring isomorphic
embeddings are returned.

The bottom-up exploration is brute-force in nature and explores all possible combinations. However, it offers some concrete directions to improve.
\begin{enumerate}
\item \textbf{Seed selection:} Can the seed edge be more strategically selected so that good candidates are found early?
\item \textbf{Seed mapping:} Can the seed edge be mapped to an edge $e$ in the target graph more intelligently so that with a high likelihood an isomorphic embedding is found around $e$, which is also highly similar to query graph $q$?
\item \textbf{Pruning bad candidates:} Can a candidate subgraph be pruned early by guaranteeing that it cannot be in the top-$k$?
\end{enumerate}

%Thm.~\ref{thm:np_hard} establishes that top-$k$ relationship-aware querying is NP-hard. 
\subsection{Seed Selection and Mapping}

Intuitively, we should explore the mapping that has the highest chance of
growing into high-scoring isomorphic match to the query. An obvious
approach is to compute edge similarity (Eq.~\ref{eq:edgesim}) between all pairs
of query edges and target graph edges, and choose the most similar pair. The
computation cost for this is $O(|E_q| \cdot |E|)$. Since $|E|$ is a
large number, we need a more efficient algorithm for this task. The similarity
between two edges is essentially a weighted min-max similarity between two
high-dimensional points, since each edge is represented by a $d$-dimensional
relationship vector. Indexing high-dimensional points have been extensively
studied \cite{indexingbook}, and we choose R-tree \cite{rtree} for our
framework. 
R-tree recursively partitions the entire dataset of relationship vectors into \emph{MBRs}. 

\begin{figure}[t] % [!ht]
\centering
\includegraphics[width=3.0in]{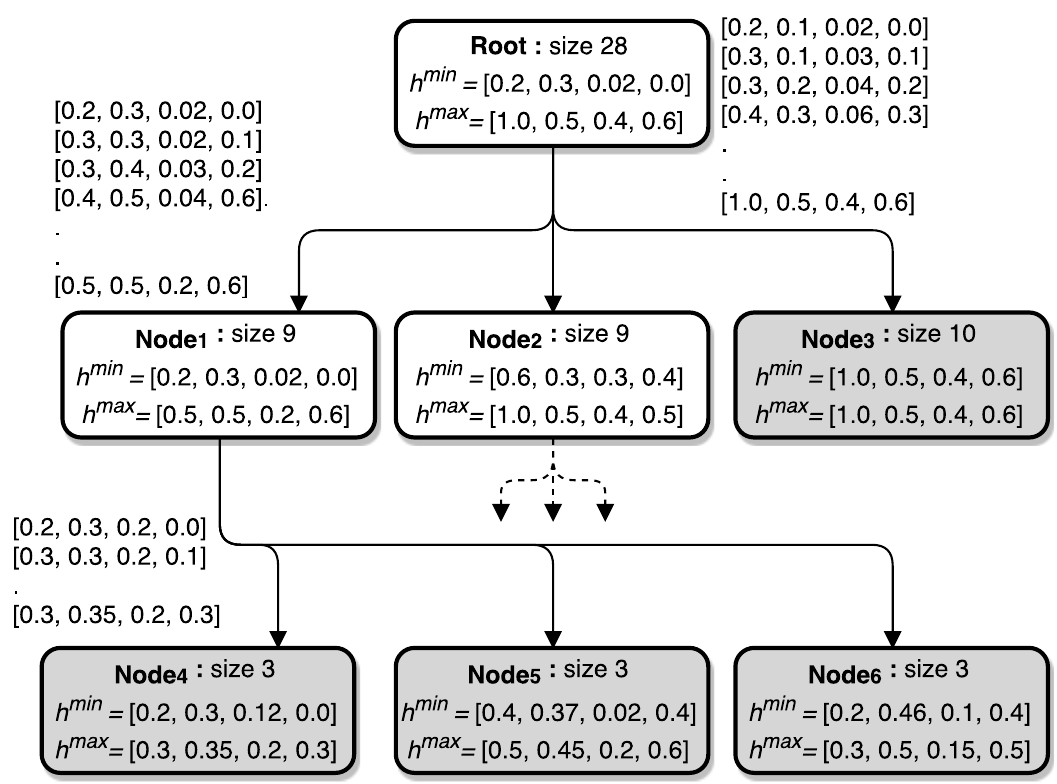}
\figcaption{\textbf{The structure of the R-tree. The leaf nodes are colored in grey. We do not explicitly show the relationship vectors in the leaf nodes due to space limitations.}}
\label{fig:CGQ_Tree}
\vspace{-0.10in}
\end{figure}

\begin{defn}[Minimum Bounding Rectangle, MBR]
	Given a set of $d$-dimensional relationship vectors, $\mathbb{D} = \{s(e_1),\cdots,$ $s(e_m)\}$, an MBR $\mathcal{H}$ on $\mathbb{D}$ is the smallest
	axis-parallel $d$-dimensional hyper-rectangle that encloses all vectors in
	$\mathbb{D}$.
\end{defn}

% Since R-tree is a well-known algorithm, the procedure to recursively partition the dataset into MBRs is provided in Appendix~\ref{app:rtree}. 

An MBR $\mathcal{H}$ can be uniquely represented by the co-ordinates of its \emph{smallest} and \emph{largest} diagonal points $h^{min}$ and $h^{max}$ respectively. More specifically, $h^{min}= [h_1^{min},\ldots ,h_d^{min}]$, where $h_i^{min}= \min_{\forall s(e) \in\mathbb{D}} \{s(f_i,e)\}$, and
%Recall from Def.\ref{def:rv} that $s(f_i,e)$ denotes the $i^\text{th}$ dimension in the relationship vector $s(e)$.
%  h_i^{max} &= \max_{\forall s(e)\in\mathbb{D}} \{s(f_i,e)\}.
$h^{max}$ is defined analogously.
An example of an R-tree is shown in Fig.~\ref{fig:CGQ_Tree}.

We use the notation $e \in \mathcal{H}$ to denote that the relationship vector $s(e)$ of $e$ is contained within MBR $\mathcal{H}$. Mathematically, this means $\forall i,\; h_i^{min}\leq s(f_i,e)\leq h_i^{max}$. We define the similarity between a relationship vector $s(e_q)$ of an edge $e_q$ and an MBR $\mathcal{H}$ as follows:
{\small\begin{align*}
  \label{eq:mbrsim}
  eSim(e_q,\mathcal{H}) &= \max_{\forall e \in \mathcal{H}} \{eSim(e_q,e)\}\\
	%&= \max_{\forall e \in \mathcal{H}} \sum_{i=1}^{d}w(f_i,e_q) \Gamma (s(f_i,e_q), s(f_i,e)), \\
	&= \sum_{i=1}^{d}w(f_i,e_q) \cdot \max_{s(f_i,e) \in [h_i^{min},h_i^{max}]} \left\{\Gamma (s(f_i,e_q), s(f_i,e))\right\} \\
\end{align*}}
\vspace{-0.20in}

Simply put, $eSim(e,\mathcal{H})$ provides the maximum similarity between $e$ and any possible edge contained within $\mathcal{H}$. This similarity value can be used to provide the following upper bound on $\mathbb{D}$.
%Iterating through all $e'$ however is not feasibly since $\mathcal{H}$ contains infinitely many vectors. Fortunately, we can compute $eSim(e,\mathcal{H})$ from just the diagonals co-ordinates of $\mathcal{H}$. 
%can be employed to obtain an upper bound on the similarity between $e$ any edge $e'$ contained within $\mathcal{H}$. More formally, we can state the following. 
\begin{thm}
  \label{thm:ub}
  If $\mathcal{H}$ is an MBR on a set of $d$-dimensional relationship vectors $\mathbb{D}$ and $e_q$ is a query edge then,
\begin{align*}
  \max_{\forall s(e)\in\mathbb{D}} \{eSim(e_q,e)\} \leq eSim(e_q,\mathcal{H}).
\end{align*}
\end{thm}

We use Theorem~\ref{thm:ub} 
%allows us to upper bound the maximum similarity between a query edge and a set of edges in the target graph. Consequently, we can 
to search for the most similar edge to the query edge $e_q$ using the \emph{best-first} search. Specifically, 
% R-tree allows us to quickly search edges that are similar to a query edge. Intuitively, we need to follow a branch that is likely to lead to the best matching edge. Thus, 
starting from the root node, we prioritize all child nodes based on the distance of their MBRs to the query edge. The best child node is then chosen to explore next and the process is recursively applied till we reach a leaf node (MBR without a child). Once a leaf node is reached, we compute the similarity to all nodes in this MBR and retain the highest scoring one. The \emph{best-first} search procedure stops when all MBRs that remain to be explored have a maximum possible similarity smaller than the highest scoring target edge we have found till now. We discuss this more formally in Sec.~\ref{sec:query}.
%We next build on this platform to develop a tree that facilitates rapid pruning of the search space.

\begin{comment}
However, R-tree (or its variants \cite{r*tree}) makes several assumptions that do not hold in our problem. Thus, we cannot use an R-tree as is, although our index
structure is inspired from its design. Similar to R-trees, we use the concept
of \emph{minimum bounding rectangles (MBR)} to summarize the relationship
vectors of a set of edges. The key aspects in which RAQ differs from R-tree are as follows.
\begin{itemize*}
  \item The distance function is assumed to be Euclidean. In our problem, the distance
between two edges is computed using the weighted min-max similarity defined in
Eq.~\ref{eq:edgesim}. 
  \item The computation of minimum and maximum similarity to an MBR is different in our problem due to the change in similarity function.
  \item Ours is a graph problem and thus we need to incorporate structural aspects in our search algorithm.
  \end{itemize*}
\end{comment}

\subsection{Avoiding Local Maxima}
\label{sec:cgq_heuristics}

While the above algorithm is efficient in locating a similar edge, it is prone to getting stuck in a local maxima. 
%We illustrate this issue with Fig.~\ref{fig:wv}. In this figure, each node is characterized by only one feature, which is its color. Suppose we want to find the best matching edge to $e_q$ of query graph $q$ in the target graph $G$. Consider $e_1$ and $e_2$ in the target graph. Both these edges are good matches since they preserve the color of their endpoints and therefore encode the same relationship. Recall that in relationship-aware similarity, node values are not matched; rather, we match features conserved in an edge. At this juncture, notice that $e_2$'s neighborhood is significantly different from the neighborhood of $e_q$. More specifically, the structural neighborhood of $e_2$ is much different. Furthermore, while $e_q$'s neighborhood is homogeneous in terms of color, $e_2$'s neighborhood is a mixture of different colors. Consequently, even though $e_2$ is a good match to $e_q$, it would be hard to find good matches for the neighboring edges of $e_q$ with those in the neighborhood of $e_2$. As a result, matching $e_q$ with $e_2$ is unlikely to lead to a graph with high relationship-aware similarity. In contrast, $e_1$ is not only similar to $e_q$, but is also situated in a similar neighborhood with homogeneous colors. Consequently, the relationship encoded by edges in $e_q$'s neighborhood are likely to be similar to the relationships encoded by edges in $e_1$'s neighborhood. Therefore, $e_1$ should be preferred over $e_2$. 
 % To summarize the above discussion,
Specifically, searching for similar edges may lead us to a good match in a bad neighborhood (local maxima), and such leads should be avoided. 
%A neighborhood is bad if the distribution of features is different from the neighborhood around the query edge. Therefore, an important question arises.
A neighborhood is bad if one or both of the following are true:
\begin{enumerate}
   \item The structure around the mapped edges is different.
   \item The distributions of relationships around the mapped edges are different and, therefore, even if an isomorphic mapping is found, the similarity among the mapped edges (relationship vectors) is likely to be low.
\end{enumerate}
%There is however one weakness of neighborhood vectors; they do not distinguish between important providing context and those that are irrelevant for the given query. This weakness can be addressed if we instead match the weight vectors of two edges. Weight vectors takes neighborhood vectors as input and computes the statistical significance of each feature in the neighborhood vector. 
\textit{Can we avoid dissimilar neighborhoods without explicitly performing isomorphism tests?} 

%\subsubsection{Neighborhood Signature:} 
%\label{sec:wv}
\noindent\textbf{Neighborhood Signature:} 
We answer the above question by mapping the neighborhood of an edge into a feature vector such that if two neighborhoods are similar, then their feature vectors are similar as well. This is achieved using \emph{Random Walks with Restarts (RWR)}.
%. Towards that goal, we can construct a mean relationship vector of all edges within $\delta$-hop neighborhood of the query edge $e_q$ and its mapped edge $e$. The mean relationship vector is the resultant vector by taking the mean of each dimension of all relationship vectors within $\delta$-hop neighborhood. It is easy to see that if two neighborhoods have similar relationship vectors, then their means would be similar as well. Certainly, there could be false positives; i.e., dissimilar neighborhoods with same mean relationship vector. In addition, this approach looks only into distribution of relationships and ignores the structural aspect. To better reflect the structure, instead of computing the mean relationship vector within $\delta$-hop neighborhood, we perform \emph{random walk with restarts (RWR)}. 
RWR starts from a source node and iteratively jumps from the current node to one of its neighbors with equal probability.
%Each neighbor has an equal probability of becoming the new
%station of the walker. 
To bound the walker within a $\delta$-hop neighborhood with a high likelihood, RWR employs a restart probability of $\alpha=\frac{1}{\delta}$. At any given state, with $\alpha$ probability, the walker returns to the source. Otherwise, the walker jumps to a neighbor.

Let $e$ be the edge whose neighborhood needs to be captured.
We arbitrarily choose one of the endpoints of $e$ as the source node.
RWR is initiated from it and for each edge $e'$ that is traversed,
a
\emph{counter} $c(e')$ records the number of times
$e'$ have been traversed. The RWR is performed for a large number of iterations
%continues till the \emph{distribution} of
%counters converges, i.e., $\frac{c(e')}{C}$, where $C$ is the total number of
%jumps taken till now. After convergence, 
 after which, we compute the
\emph{neighborhood signature} of edge $e$ as
a $d$-dimensional feature vector $N(e)=[n_1,\cdots,n_d]$, where 
\begin{equation}
\label{eq:signature}
n_i=\sum_{\forall e' \in \mathbb{E}} \frac{c(e')}{C}\times s(f_i,e')
\end{equation}
where $\mathbb{E}$ is the set of all edges visited in the RWR and $s$ is as defined in Def.~2 and $C$ is the total number of iterations.
%jumps taken till now..

\begin{comment}
\begin{figure}[t] % [!ht]
\centering
\includegraphics[width=2.7in]{pdf/wv.pdf}
\figcaption{\textbf{The importance of neighborhood structure and feature distribution in RAQ.}}
\label{fig:wv}
\end{figure}
\end{comment}

%More simply, instead of taking the mean, we take a weighted aggregate of the relationship vectors in the neighborhood. 
%The weight of a relationship vector is proportional to the number of times the corresponding edge is visited in the RWR. 
The signature captures the distribution of relationships around the
source edge $e$. Edges that are closer to $e$ have a higher say.
Consequently, the signature not only captures the relationship vectors in the
neighborhood, but is also structure sensitive. While it is still possible for
two dissimilar neighborhoods to have similar signatures, the chances that two
dissimilar neighborhoods produce similar signatures across all $d$ dimensions is
small.

We construct the neighborhood signature of all edges in the target graph as part
of the index construction procedure. The signatures of the edges in the query
graph are computed at query time.
The neighborhood hop parameter $\delta$ is typically a
small value since query graphs for knowledge discovery tasks are normally
small\cite{exemplar}. In our empirical evaluation, we set $\delta$ to $3$.

\begin{comment}
\textbf{Time Complexity: } The weight vector calculations are done for each edge and it requires exploration of the $h$ hop neighborhood around an edge. If $\eta$ is the average degree in the target graph, the time complexity is $O(d \eta^h)$ on average for a single edge. $h$ is typically 1 or 2. Sorting across $d$ dimensions requires $O(d|E|log|E|)$ cost. After incorporating the cost of building R-tree, the total index construction of CGQ is $O(d|E|log|E|+d \eta^h |E|)$.
%After weight calculations, we sort the edges in homogeneous leaf nodes (node containing only one type of $s(e')$), so if the number of homogeneous leaf nodes is $l$ and the maximum size of the node is $\alpha$, the time complexity of sorting is $O(l \alpha \log \alpha)$. 

\textbf{Space Complexity: } Weight vectors incur a storage of $O (d|E|)$ cost. The overall cost of the entire CGQ index is also $O(d|E|)$.
%, for storing $d$ sorted lists in each of the $l$ bins.

The incorporation of the weight vectors completes the CGQ index. In the next section, we discuss how the R-tree in conjunction with the sorted weigh vectors are employed to answer relationship-aware top-$k$ queries.
\end{comment}

\subsection{RAQ Search Algorithm}
\label{sec:query}

\begin{comment}
We start from a common 1-edge subgraph, and grow them through neighboring edges to form an MCS. The initial 1-edge subgraph is formed by mapping an arbitrary edge in the query graph to an arbitrary edge in the target graph. With the help of RAQ index structure, the goals of our searching algorithm are two-fold. 
\begin{enumerate*}
  \item \textbf{Detect promising leads: }Instead of mapping two arbitrary edges, identify a promising mapping such that the resultant 1-edge common subgraph leads to an MCS with high relationship-aware similarity.
  \item \textbf{Prune early: }Instead of discovering that the relationship-aware similarity of an MCS is low after fully growing it, prune them early in their initial stages of formation.
  \end{enumerate*}

  With these objectives in mind, we design our searching algorithm. 
\end{comment}

\begin{figure}[t] % [!ht]
\centering
\includegraphics[width=3.3in]{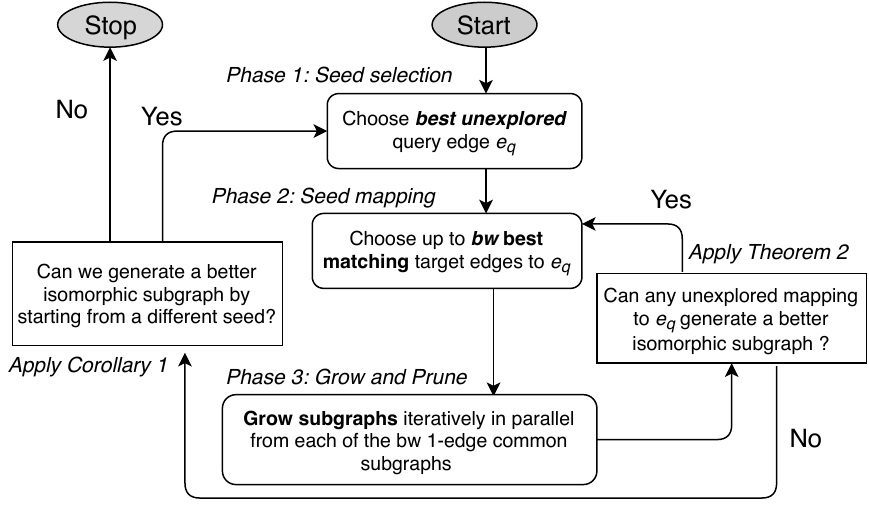}
\vspace{-0.20in}
\figcaption{\textbf{Flowchart of the searching algorithm in RAQ.}}
\label{fig:flowchart}
\end{figure}

Fig.~\ref{fig:flowchart} outlines the flowchart and Alg.~\ref{algo:cgq} presents
the pseudocode. There are three major phases: selecting the most promising query
edge $e_q$ for exploration (\emph{seed selection phase}), identifying the best matching
target edges (\emph{seed mapping phase}), and growing subgraphs from these initial
seeds in bottom-up manner (\emph{growth phase}).

Before we execute any of the phases, two operations are performed. First, we
compute the neighborhood signatures of each edge in the query (line 1). Second,
we use a priority queue called $Ans$, which stores the $k$ most similar
subgraphs identified till now (line 2). Initially, $Ans$ is empty.
  
\begin{figure*}[t]
\centering
\includegraphics[width=6.2in]{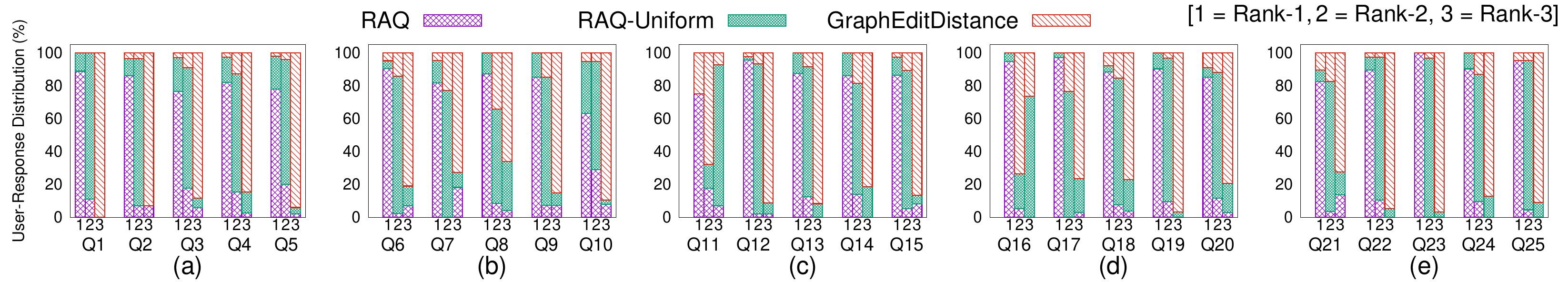}
\vspace{-0.10in}
\figcaption{\textbf{The distribution of the received user responses.}}
\label{fig:user_study_response_distribution}
\end{figure*}
\subsubsection{Phase 1: Seed Selection}
We select the unexplored query edge $e_q\in E_q$ that has the highest similarity to the root node of R-tree (line 5). An edge is \emph{unexplored} if it has not already been selected in this phase in some earlier iteration.

\subsubsection{Phase 2: Seed Mapping}
\label{sec:phase2}
We use \emph{best-first search} to find the leaf node in R-tree that has the highest similarity to $e_q$. 
%Specifically, we initialize a priority queue, \textit{Cands} (line 6). $Cands$ stores MBRs corresponding to nodes in R-tree and orders them in descending order of their similarity to the query edge $e_q$ (Theorem~\ref{thm:ub}). Initially, $Cands$ contains the root node. We iteratively pop the top node (or MBR) from $Cands$ and insert all its children to $Cands$ till a leaf node is reached. 
In this leaf node, we perform a \emph{beam stack search}\cite{beamStackSearch} with \emph{beam width} $bw$ (lines 13-20). Instead of mapping $e_q$ to the most similar target edge, beam stack search selects the $bw$ target edges that have the highest neighborhood similarity to $e_q$ and explores them further. %This strategy is motivated by the property that a target edge is a good match to $e_q$ if it has high relationship-aware similarity and it is located in a similar neighborhood. Since the selected leaf node has the highest similarity to $e_q$ among all unexplored leaf nodes in R-tree, it is likely that all target edges in this node are similar to $e_q$. Thus, we shift our focus to neighborhood similarity. 
The \emph{neighborhood similarity} between two edges $e_q$ and $e$ is the weighted $L_1$ distance between their neighborhood signatures.  
\begin{equation}
\nonumber
\text{Neighborhood\_Similarity}(e_q,e)=\sum_{i}^d w_i \cdot \|N(i,e_q)-N(i,e)\|
\end{equation}
Recall that $w_i$ is the importance of the $i^\text{th}$ relation and more important relations have a higher say in defining the neighborhood similarity.
 
\subsubsection{Phase 3: Grow and Prune}
\label{sec:phase3}
Each of the $bw$ edges selected in the previous phase generates a common 1-edge subgraph of the query and the target graph. We now grow them one edge at time to larger subgraphs (line 19). On any such subgraph $g=(V_g,E_g)$ we can apply the following bound.
\begin{thm}
  \label{thm:mcsub}
  The maximum similarity of any isomorphic match to query graph $q$ formed by growing $g$ further is $\widehat{RAQ}(q,$ $g)=RAQ(q,g)+|E_q|-|E_g|$.
\end{thm}
\begin{proof} The maximum similarity between any two edges is $1$ (Eq.~\ref{eq:edgesim}). Therefore, the maximum similarity contributed from the edges yet to be added is at most $|E_q|-|E_g|$.
	The theorem follows.
	%Thus, $RAQ(q,g)\leq \widehat{RAQ}(q,g)$.%=RAQ(q,g)+|E_q|-|E_g|$. 
\end{proof}

\begin{corl}
  \label{cor:mcsub}
  Let $e_q$ be a query edge and $\mathcal{H}$ be an MBR in the R-tree. The maximum similarity of any isomorphic subgraph formed by mapping $e_q$ to any edge contained within $\mathcal{H}$ is $\widehat{RAQ}(e_q,\mathcal{H})=eSim(e_q,\mathcal{H})+|E_q|-1$.
\end{corl}
\begin{algorithm}[t]
  \caption{RAQ Search Algorithm}
  \label{algo:cgq}
  {\scriptsize
\begin{algorithmic} [1]
\REQUIRE $G=(V,E,\mathbb{F})$, $q(V_q,E_q,\mathbb{F}_q),\,k, RAQ Index$
\ENSURE Top-$k$ subgraphs of $G$ with highest relationship-aware similarity to $q$.
\STATE Compute neighborhood signature for each edge in $q$
\STATE $Ans\leftarrow \emptyset$   \COMMENT{Priority queue of maximum size $k$. Stores isomorphic matches in descending order of similarity.}
\STATE $Cands\leftarrow \emptyset$ \ignore {Priority Queue to store MBRs}
\WHILE {$\exists e\in E_q,\;\widehat{RAQ}(e,rootNode.MBR)>Ans.leastValue()$ and $e$ has not been explored }
\STATE $e_q \leftarrow$ unexplored query edge with highest similarity to MBR in root node
\STATE Initialize $Cands$ with the root node of R-tree
\WHILE {$Cands$ is not empty}
\STATE $n \leftarrow Cands.poll()$ \ignore {Most similar node (MBR) to $e_q$}
\STATE $max \leftarrow \widehat{RAQ}(e_q,n.MBR)$ \ignore{Corollary~\ref{thm:mcsub}}.
\IF {$max \leq \allowbreak Ans.leastValue()$}
\STATE \textbf{break} \ignore {answer set computation terminates}
\ENDIF
\IF {$n$ is a leaf node}
\WHILE {$\widehat{RAQ}(e_q,n.MBR)>Ans.leastValue()$ and $n$ contains unexplored edges}
\STATE $PQ\leftarrow$ Priority queue containing $bw$ unexplored edges with similar neighborhoods to $e_q$  \ignore{Phase 2 (Sec.~\ref{sec:phase2}). }
\WHILE{$PQ$ is not empty}
\STATE $g\leftarrow PQ.poll()$ \ignore{ subgraph $g$ that maximizes $\widehat{RAQ}(g,q)$}
\IF {$\widehat{RAQ}(g,q) \leq \allowbreak Ans.leastValue()$}
\STATE \textbf{break} \ignore {computation ends for this batch of $bw$ edges}
\ENDIF
\STATE $\mathbb{G}\leftarrow$ Grow $g$ through all possible 1-edge extensions to form common subgraphs
\STATE Add all isomorphic matches in $\mathbb{G}$ to $Ans$ and remaining subgraphs in $\mathbb{G}$ to $PQ$
\ENDWHILE
\ENDWHILE
\ELSE
\STATE Add all children of $n$ to $Cands$.
\ENDIF
\ENDWHILE
\ENDWHILE
\RETURN $Ans$
\end{algorithmic}}
\end{algorithm}

Equipped with the above upper bounds, we prioritize each of the initial 1-edge
common subgraphs based on $\widehat{RAQ}(q,g)$ (Theorem~\ref{thm:mcsub}). More
specifically, we initialize another priority queue $PQ$ and insert all 1-edge
subgraphs in $PQ$ (line 14). The subgraph $g$ with the highest upper bound is
popped from $PQ$. We check if $\widehat{RAQ}(q,g)$ is larger than the
$k^\text{th}$ highest similarity score in $Ans$ (line 17). If yes, we
explore all possible single edge extensions of $g$ to create common subgraphs
with one more edge (line 20). If $g$ becomes isomorphic to $q$ after extension,
then we add it to $Ans$. Otherwise, we insert it back to $PQ$ (line 20).
Otherwise,
if the check (at line 17) fails, then we are guaranteed that none of the
unexplored subgraphs in $PQ$ can lead to a better solution than what we already
have and, hence, Phase 3 completes (lines 7-18).

\subsection{Properties of RAQ Framework}

\textbf{Correctness Guarantee:} The RAQ framework provides the \emph{optimal answer set}. We do not prune off any possibility without applying Theorem~\ref{thm:mcsub} or Corollary~\ref{cor:mcsub}. Consequently, we do not lose out on any candidate that can be in the top-$k$ answer set. 

\textbf{Index construction cost:} The cost of constructing the R-tree is $O(d|E|\log (|E|))$. 
%A detailed explanation for this cost is provided in Appendix~\ref{app:rtree}. 
We also compute and store the neighborhood signature of each edge through random walk with restarts. This step can be performed in $O(|V|^2)$ time for the entire graph\cite{rwr,rwr1}. Thus, the total computation cost is $O(|V|^2+d|E|\log (|E|))$.

\textbf{Memory Footprint:} R-tree consumes $O(d|E|)$ space since it stores $d$-dimensional relationship vectors of all edges. The neighborhood signatures are also $d$-dimensional feature vectors and hence the total cost is bounded by $O(d|E|)$.

\textbf{Querying Time:} As stated earlier, the problem is NP-complete. Hence, akin to most existing graph search algorithms\cite{NEMA, exemplar, ctree}, the worst case running time remains exponential. However, with the application of efficient heuristics and effective pruning bounds, we compute the top-$k$ answer set in milliseconds as elaborated next.

\section{Experiments}
\label{sec:exp}

\begin{table}[b]
  \tabcaption{\textbf{Summary of datasets used in our experiments.}}
\centering
%\scalebox{0.7}{
\scalebox{0.70}{
  %\begin{tabular}{ |>{\centering}m{2cm}<{\centering}|>{\centering}m{1.1cm}<{\centering}|>{\centering}m{1.2cm}<{\centering}|>{\centering}m{1cm}<{\centering}|>{\centering}m{1.2cm}<{\centering}| } 
  \begin{tabular}{lrrlrr}%{|c|c|c|c|c|c|}
    \hline
    %\textbf{Dataset} & \textbf{Number of Nodes} & \textbf{Number of Edges} & \textbf{Mean Degree} & \textbf{Number of Features} \tabularnewline
    \textbf{Dataset} & \textbf{\# Nodes} & \textbf{\# Edges} & \textbf{Type} & \textbf{Mean Degree} & \textbf{\# Features} \tabularnewline
    \hline 
%    Northeast &  1.4K & 15K & Undirected & 10.50 & 5\tabularnewline
    IMDb & 88K  & 781K & Undirected & 17.59 &7 \tabularnewline
    DBLP & 3.2M & 6.8M & Directed & 2.13 & 5\tabularnewline
    DBLP (co-author) & 1.8M & 7.4M & Undirected & 8.44 & 5\tabularnewline
    Pokec & 1.6M & 30.6M & Directed & 19.13 & 6 \tabularnewline
    \hline
  \end{tabular}}
  \label{tab:dataset} 
\end{table}

In this section, we evaluate the RAQ paradigm through user surveys, case studies, and scalability experiments.

\begin{comment}
and establish the following:
\begin{itemize}
\item \textbf{Quality: }Relationship-aware similarity extracts meaningful results that traditional graph similarity functions such as edit distance are unable to identify.
\item \textbf{Scalability: } RAQ scales to large graphs containing millions of nodes and edges.
\end{itemize}
\end{comment}

\subsection{Datasets}
\label{sec:datasets}

We consider a mix of various real (large) graphs (Table~\ref{tab:dataset}).

\begin{comment}
$\bullet$ The \textbf{Northeast (NE)} biodiversity dataset \cite{ne,graphchi}
		comprises of $1434$ spatial points, with each node possessing $5$ types
		of information (features): (i) Bio-diversity richness index, (ii)
		Disturbance index, (iii) Medicinal property, (iv) Economical property,
		and (v) Forest type. If two nodes are spatially close, they are
		connected by an edge.  Forest type is a categorical feature possessing
		27 unique categories, while all other features are numerical.
\end{comment}

$\bullet$ The \textbf{IMDb} dataset \cite{imdb} contains 
%information like cast, trivia, user ratings etc. about movies, television series, internet streams etc. Similar to the DBLP co-authorship network, we construct 
a collaboration network of movie artists, i.e., actors, directors, and producers. Each node is an artist featuring in a movie, and two artists are connected by an edge if they have featured together in at least one movie. Each node possesses $4$ features: (i) year of birth, (iii) the set of all the movies the artist has featured in, (iv) the most prevalent genre among his/her movies, and (v) the median rating of the movie set. %The median rating and the year of birth are numerical features, while the rest are categorical. (i) the primary profession (ex: actor), (ii) the primary industry (ex: Hollywood), (iii) year of birth, (iv) the vital status (alive/dead), (v) the set of all the movies the artist has featured in, (vi) the most prevalent genre among his/her movies, and (vii) the median rating of the movie set.
%We use the IMDb dataset to conduct a user survey for assessing the quality of the answer sets returned by RAQ and baseline techniques. During our preliminary investigations, we realized that many users were not familiar with older artists and hence, were unable to judge the quality of an answer. Thus, we removed all the artists born before $1948$.

$\bullet$ The \textbf{DBLP} dataset \cite{dblp} represents the citation network
		of research papers published in computer science. Each
		node is a paper and a directed edge from $u$ to $v$ denotes
		paper $u$ being cited by paper $v$. Each node possesses $5$ features: (i) publication venue,
		(ii) the set of authors, (iii) year of
		publication, (iv) the rank of the venue, and (v) the subject area of the venue. The publication year and venue rank are numerical
		features while the rest are categorical in nature. The first three
		features are obtained from the DBLP database while the last two are
		added from the CORE rankings portal \cite{core_ranking}.
		%categorical features: venue, authors[ids, hashset], (numerical
		%features)year of publication (DBLP). (numerical features)CORE dataset:
		%rank of venue, A*, A, B, C, Unranked (1-5), Corresponding Subjects:
		%[hashset[] (Jaccard).

$\bullet$ The \textbf{DBLP (co-author)} dataset \cite{dblp} represents the
		co-authorship network of the research papers. Each node is an author of a paper and edges denote
		collaboration between two authors. Each node possesses $5$ features: (i) number of papers published by the author,
		(ii) the number of years for which the author has been active
		(calculated as the difference between the latest and the first paper),
		(iii) the set of venues in which the author has published, (iv) the set
		of subject area(s) of the venues in which the author has published, and
		(v) the median rank of the venue (computed as the median of the ranks
		of all the venues where the author has published). The number of papers
		published, number of years active, and the median rank are numerical
		features, while the rest are categorical. %Similar to the DBLP dataset,
%		the first three features are from the DBLP database while
%		the last two are from the CORE rankings portal \cite{core_ranking}.
		%Features: \#papers, years active (year of last paper - first), venues
		%(hashset) (DBLP), subjects (top-80 subjects, doesn't matter overall)
		%(CORE), median rank (from each venue). (5 features.)

$\bullet$  The \textbf{Pokec} dataset \cite{snap} presents data collected from
		an on-line social network in Slovakia. Each node is a user of the
		social network and a directed edge $e=(u,v)$ denotes
		that $u$ has a friend $v$ (but not necessarily vice versa). %The dataset
		%originally contains $59$ node features. However, since every node does
		%not possess values for all the $59$ features, we choose the $6$ most
		%frequent 
		Each node is characterized by six features: (i) gender, (ii) age, (iii) joining date, (iv) timestamp of last login (v) public: a boolean value indicating whether the
		friendship information of a user is public, and (vi) profile
		completion percentage. While public and gender are
		categorical features, the rest are numerical.
		%1-8 (user-id:1 and region: are not being used.), Last login: list of
		%all last logins. Randomly picking one, Age:, Gender: Checking
		%bit-bucket
%\noindent
%

\vspace{-0.30in}
\subsection{Experimental Setup}
All experiments are performed using codes written in Java on an Intel(R)
Xeon(R) E5-2609 8-core machine with 2.5 GHz CPU and 512 GB RAM running Linux
Ubuntu 14.04.

\textbf{Query graph:} Query graphs are generated by selecting random subgraphs
from the target graph. Following convention in graph querying \cite{NEMA,exemplar}, we considered
queries with size up to $10$ edges. All reported results
were averaged over $30$ random query graphs.
%The graph is grown by selecting edges uniformly at random from a
%multiset containing all the edges connected to the current graph. Since our
%algorithm and the problem formulation is edge oriented, we define the size of a
%graph as the number of edges.

%Next, we state the methods, their parameters and the evaluation metrics used
%for obtaining the results.

\begin{figure}[t]
  \begin{tabular}{ccc}
	\subfloat[Weight Correlation \label{fig:weight_survey1}]{\includegraphics[width=0.99in]{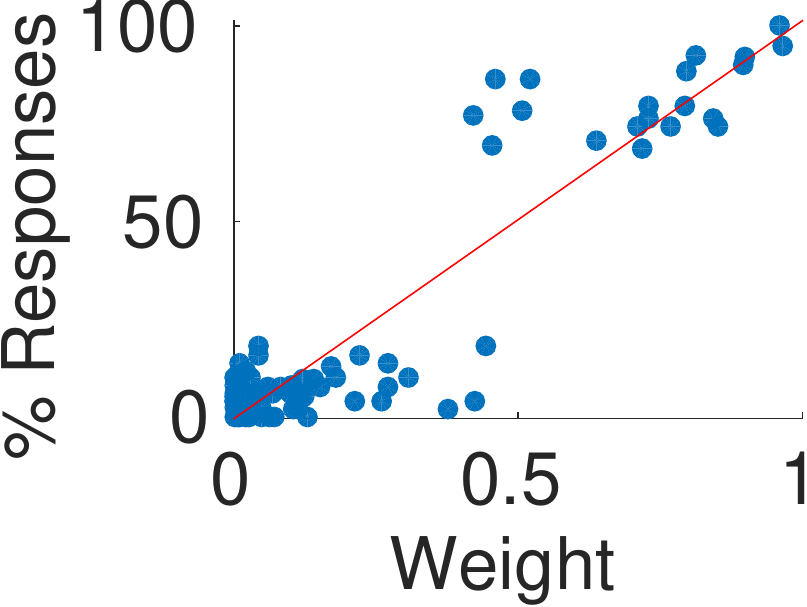}} &
    \subfloat[Relation Weights \label{fig:fweight_survey1}]{\includegraphics[width=0.99in]{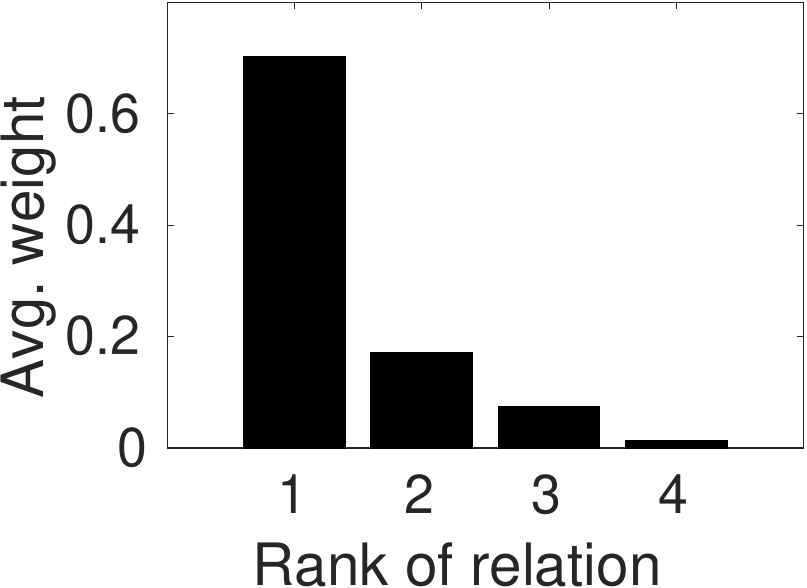}} &
    \subfloat[Rank Correlation\label{fig:rho_survey2}]{\includegraphics[width=0.99in]{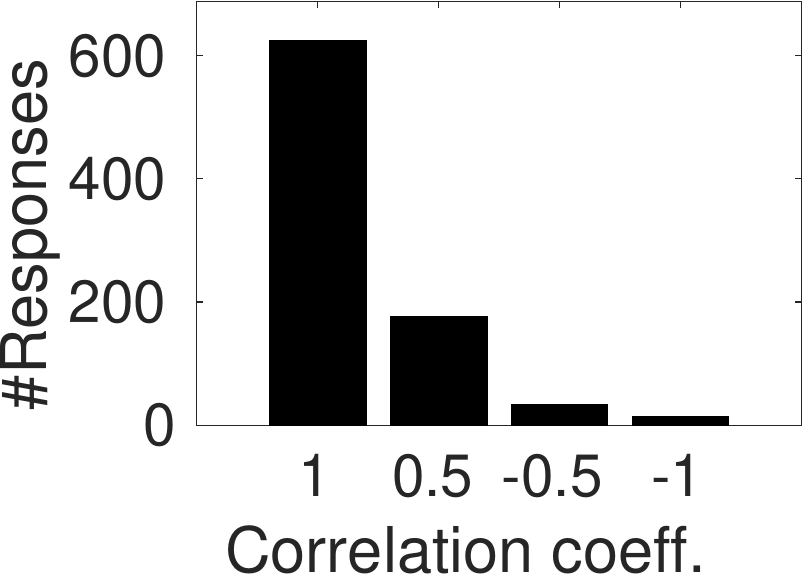}}%&
%    \subfloat[ \label{fig:BaselineBFS_Comparison_NE}]{\includegraphics[width=1.30in]{pdf/BaselineBFS_Comparison_NE_Bar.pdf}} &
  %  \subfloat[\label{fig:BranchingFactor}]{\includegraphics[width=\imageWidthIcePlot]{pdf/BranchingFactor.pdf}} %\\  
  \end{tabular}
\vspace{-0.20in}
  \figcaption{\textbf{(a) Correlation between weights and user-perceived relationship importance. Variation in (b) average weights from most to least important relationship, (c) Correlation across user responses.
}}
\end{figure}
\textbf{Baselines:} To measure the quality of graphs returned by RAQ, we compare its performance to three baselines:

$\bullet$ \textbf{RAQ-uniform:} To answer the question whether setting the
weight of a relationship in proportion to its statistical significance helps,
we implement RAQ-uniform as the setting where all relationships are considered
equally important.

$\bullet$ \textbf{Graph edit distance (GED):} The \emph{edit distance}
between graphs $q$ and $g$ is the minimum \emph{cost} of \emph{edits} required
to convert $q$ to $g$\cite{ctree}. An edit corresponds to deleting a node or an edge,
adding a node or an edge, or substituting a node. Each edit has a cost. Deletions
and additions of nodes or edges have a cost of $1$. The cost of substituting a
node $u$ with $v$ is proportional to the distance between them. We use the
\emph{min-max distance} between the feature vectors of two nodes to compute the
distance. The replacement cost $rep(u,v)$ is:
\begin{alignat}{3}
\nonumber
rep(u,v)&=1-sim(u,v)\\
\nonumber
		sim(u,v) &= \frac{1}{d} \sum_{i=1}^{d}s_i(u,v) \qquad \text{where}\\
\nonumber
	    s_i(u,v) &=
			\begin{cases}
				\frac{\min\{f_i(u),f_i(v)\}}{\max\{f_i(u),f_i(v)\}} \quad \text{if} \quad f_i \text{ is real valued} \\
				1 \quad \text{if} \quad f_i \text{ is categorical and } f_i(u) = f_i(v) \\
				0 \quad \text{if} \quad f_i \text{ is categorical and } f_i(u) \neq f_i(v)
			    \end{cases}
\end{alignat} 

$rep(u,v)$ is always within the range $[0,1]$.
%Equipped with graph edit distance, we identify the $k$ subgraphs of the target graph that are isomorphic to the query and have the lowest edit distance.

$\bullet$ \textbf{Query by example (QBE):} We choose \cite{exemplar1} to
compare against the QBE paradigm (described in Sec.~\ref{sec:qbe}). To make it
compatible to our framework,
we model the node features as edge labels.
Specifically, if $u$ and $v$ are connected in the original network, we
introduce $d$ edges between them. The $i^\text{th}$ edge between $u$ and $v$
corresponds to the $i^\text{th}$ relationship (feature), and this relationship
is captured in the form of edge label by concatenating the feature values
$f_i(u)$ and $f_i(v)$. All numerical features are discretized into bins
\cite{binning} since otherwise, there would be infinitely many edge labels.

%As discussed in Sec.~\ref{sec:intro}, none of
%the existing techniques in the literature can be used for either answering relationship-aware queries or index them. 
%an empirical
%comparison with RAQ and its querying algorithm. To this end, we incorporate the
%use of a \emph{baseline} technique, that of traditional subgraph isomorphism
%search modified to use RAQ as the similarity function. The search algorithm
%uses depth first search to find all the maximal common subgraphs. 
%Thus, the baseline technique for us is the naive algorithm discussed in XXX %Sec.~\ref{sec:naive}.
%
\noindent
\textbf{Parameters:}  The default value of $k$ in a top-$k$ query is set to $10$. 
%The restart probability used in RWRs to construct neighborhood signatures is set to $\alpha=\frac{1}{3}$. 
The branching factor in R-tree is $b=4$. %While building the R-tree, we set a node as leaf node if it contains less than $100$ edges.

%The user-base was carefully chosen to ensure \emph{diversity} in various aspects such as profession, educational background, gender, and age.
\subsection{User Survey: IMDb Dataset}
\label{sec:survey}

To assess the relevance of the results retrieved using the proposed RAQ
paradigm, we conducted two surveys spanning $170$ users each, using
collaboration patterns (subgraph queries) from the IMDb dataset. The user-base
was carefully chosen to ensure \emph{diversity} in various aspects such as
profession (engineers, researchers, homemakers, doctors, business executives
etc.), educational background (from undergraduates to doctorates in science,
engineering, medicine etc.), gender (approximately equal distribution between
females and males), and age (from 20 to 70 years old). Each user was presented
with $5$ queries randomly sampled from the master query pool, which led to a
total of $850$ user-responses for each user survey. To enable the users in
understanding the tasks better, they were briefed on the capabilities of the
RAQ framework and the semantic interpretation of a graph.
%. Moreover, using a sample input query they were introduced to the technical
%concepts of graphs (Ex: the meaning of an edge between two artists) along with
%an example of the type of relationships that may be manifested in the query.
%Both the user-surveys along with their documentation are accessible at:
%\url{https://blind-anonymous.github.io/}\footnote{Note that we do not track
%the users visiting this web page.}.

%\subsubsection{Relationship Importance}
%\label{sec:survey_weight}
\noindent\textbf{Relationship Importance:}
%As stated in Sec.~\ref{sec:relationship}, we humans value rare events, which further motivated the use of statistical significance based weighting (normalized $X^2$ values) for quantifying relationship importance. Thus, 
The first user survey studies whether statistical significance is consistent with human intuition. For this, we constructed a query pool of $20$ edges (representing collaboration between two artists) along with their node features. %Note that intuitively, a query seeking similar ``actor-director'' collaborations should always result in a collaboration between an actor and a director. Thus, to preserve the semantics between the query and the retrieved responses, the node feature: ``primary profession'' was excluded from the relationship importance analysis. With this change, each node possesses $4$ features, thereby resulting in $4$ possible relationships per edge.
For each query edge, we randomly choose two relationships (features), and the users were asked to identify the relatively more important relationship. 
%Note that both the approaches are equivalent in their capability to assist identification of relationship importance. %We also interviewed a small subset of $20$ users from our user-base, where the majority showed preference towards binary questions.
%To make this task intuitive for the users, instead of asking them to rank the $4$ relationships based on their importance, the users were presented with a binary question considering all possible relationship combinations. Specifically, given a query edge and a pair of relationships, the users were asked to identify the relatively more important relationship. Note that both the approaches are equivalent in their capability to assist identification of relationship importance. We also interviewed a small subset of $20$ users from our user-base, where the majority showed preference towards binary questions.
A large number, $692/850$ ($81.4\%$), of the received user-responses considered the relationship with higher statistical significance to be more important. Considering the null hypothesis that either relationship is equally probable to be chosen as important, this observed outcome possesses a %$X^2$ value of $335.478$, and is statistically significant with a 
$p$-value of $1\mathrm{e}{-74}$ and is, therefore, highly statistically significant.

Moving ahead, we also analyze the correlation between the weight (normalized $X^2$ value) of a relationship and the proportion of times the feature is chosen as the preferred one in the survey. Fig.~\ref{fig:weight_survey1} shows that the higher the weight of a relationship, the more likely it is to be chosen as important in the user-responses. This is also indicated by their significantly high Pearson's correlation coefficient \cite{pearson} of $0.923$. Fig.~\ref{fig:fweight_survey1} presents the average weight of the $k^\text{th}$ ranked feature, where $k$ is varied in the $x$-axis. We observe an exponential decay in weights, with the rank-$1$ relationship possessing a weight of $0.72$ on average. % (being significantly higher than rank-2,3,4 relationships) on average, and 
This result, in conjunction with Fig.~\ref{fig:weight_survey1}, shows that the highest ranked relationship has a much higher likelihood of being perceived as important in the user-responses. Overall, this survey provides substantive evidence to support the use of statistical significance in quantifying relationship importance. 

%Each user was presented with $5$ queries randomly sampled from the pool of $25$ queries, which led to a total of $850$ user-responses translating to $34$ responses per query on average.
%Note that, in addition to performing a relationship-aware match, the RAQ framework also possesses the capability to rank the relationships manifested in a query based on their importance.
%\subsubsection{Ranking Results based on Relevance}
%\label{sec:survey_raq}
\noindent\textbf{Ranking Results based on Relevance:}
The second survey evaluates the relevance of the results retrieved by RAQ and the three baselines discussed above. For this task, we considered a query pool consisting of $25$ query subgraphs (representing group-level artist collaborations) possessing a judicious mix of node features (artist types, prominence, genres, etc.). For each query, the users were presented with top-$1$ results returned by the four algorithms being studied, and were asked to rank the results based on the user-perceived relevance to the query.
To eliminate bias, the mapping of which result comes from which algorithm was masked from the user, and the order of the results presented were randomized.
For ensuring simplicity and brevity, each query was annotated with the most important relationship, which was shown to correspond well with human perceived importance in the first survey.

\begin{table}[t]
\tabcaption{{\bf Comparing the quality of RAQ with the considered baselines using both user-level and query-level responses.}}
\label{tab:user_study}
\centering
\scalebox{0.7}{
\begin{tabular}{crrrcrrr}
\hline
\bf Retrieved & \multicolumn{3}{c}{\bf User-level Responses (\%)} & & \multicolumn{3}{c}{\bf Query-level Responses (\%)} \\
\cline{2-4}
\cline{6-8}
%\cline{5-7}
\bf Rank & \bf RAQ & \bf Uniform & \bf GED & & \bf RAQ & \bf Uniform & \bf GED \\
\hline
\textbf{1} & \bf 86.0 & 10.6 & 3.4 & & \bf 100.0 & 0.0 & 0.0 \\
\textbf{2} & 10.0 & \bf 75.4 & 14.6 & & 0.0 & \bf 92.0 & 8.0 \\
\textbf{3} & 4.0 & 14.0 & \bf 82.0 & & 0.0 & 8.0 & \bf 92.0 \\
\hline
\end{tabular}
}
\end{table}

\begin{figure*}[t]
\label{fig:qualitative1}
\centering
\subfloat[Query 1 \label{fig:query1}]{\includegraphics[scale=0.62]{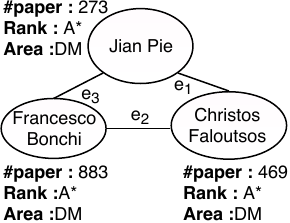}}\hfill
\subfloat[Result 1 \label{fig:result1}]{\includegraphics[scale=0.62]{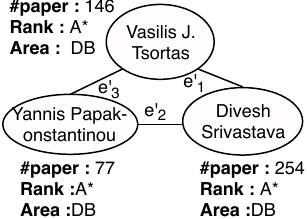}}\hfill
\subfloat[Query 2 \label{fig:query2}]{\includegraphics[scale=0.62]{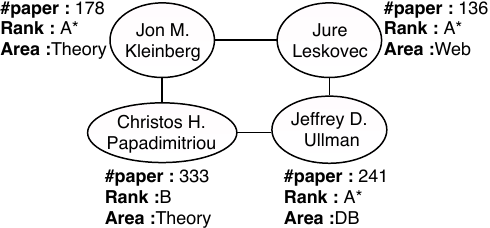}}\hfill
\subfloat[Result 2 \label{fig:result2}]{\includegraphics[scale=0.62]{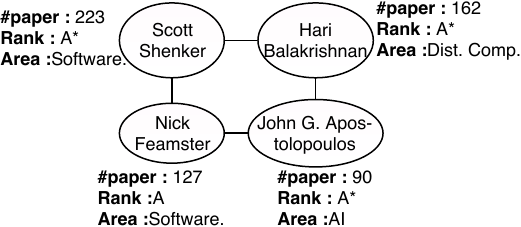}}
%\subfloat[Query 3 \label{fig:query3}]{\includegraphics[scale=0.65]{pdf/qualitative/Query_3.pdf}}\hfill
%subfloat[Result 3.1 \label{fig:result3.1}]{\includegraphics[scale=0.65]{pdf/qualitative/Result_3_1.pdf}}
%\subfloat[Result 3 \label{fig:result3.2}]{\includegraphics[scale=0.65]{pdf/qualitative/Result_3_2.pdf}}
\figcaption{\textbf{Matched subgraphs by RAQ.}}
\end{figure*}
\begin{comment}
731 90 29
85 641 124
34 119 697
\end{comment}

\begin{figure*}[t]
  \begin{tabular}{ccccc}
    \subfloat[RAQ vs GED \label{fig:QualitativeQuantitativenh}]{\includegraphics[width=\imageWidthIcePlot]{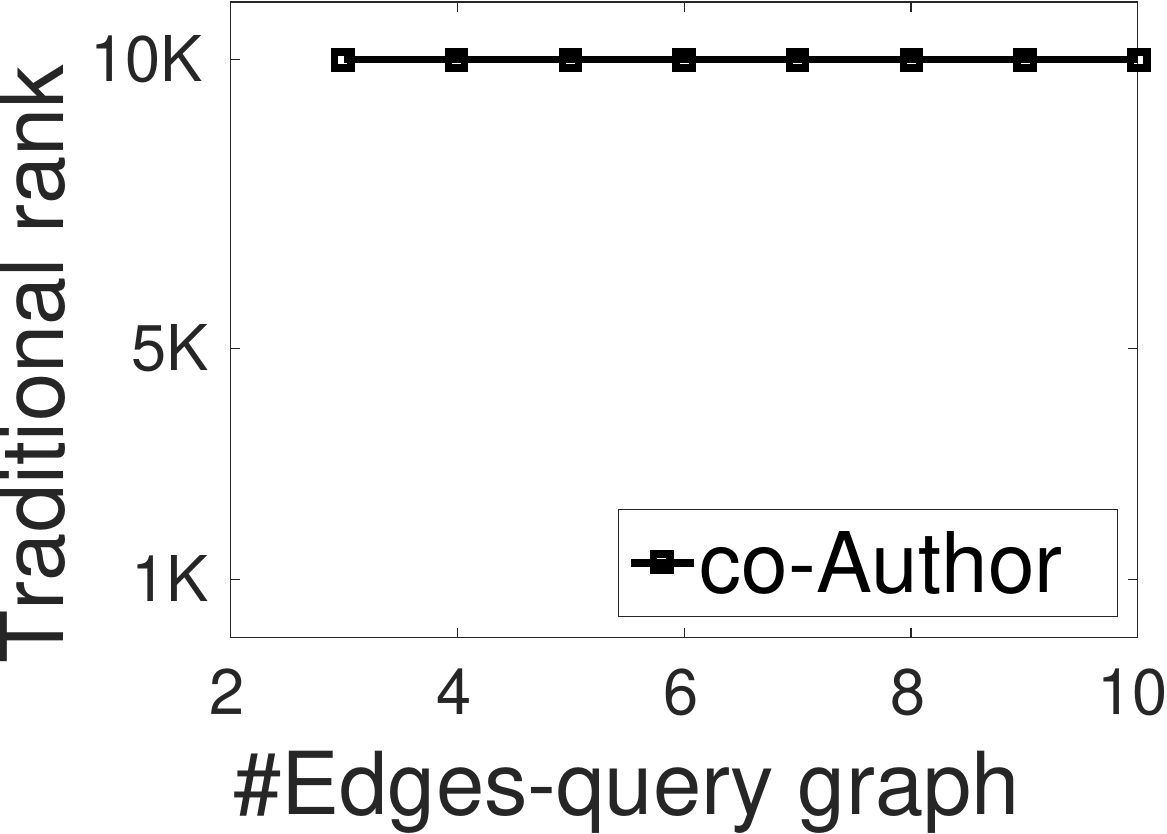}} &
	\subfloat[IMDb \label{fig:runtime}]{\includegraphics[width=1.2in]{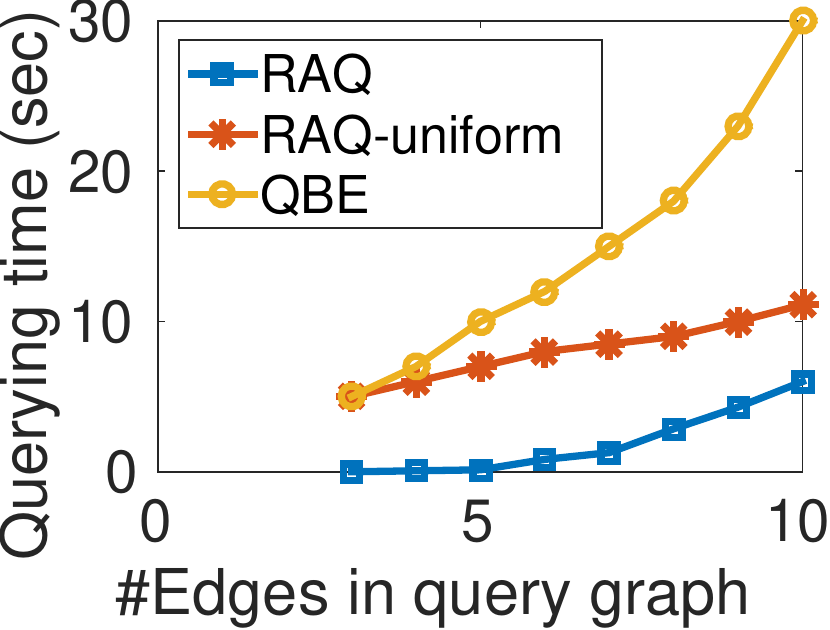}}&
    \subfloat[Large datasets\label{fig:QuerySize}]{\includegraphics[width=\imageWidthIcePlot]{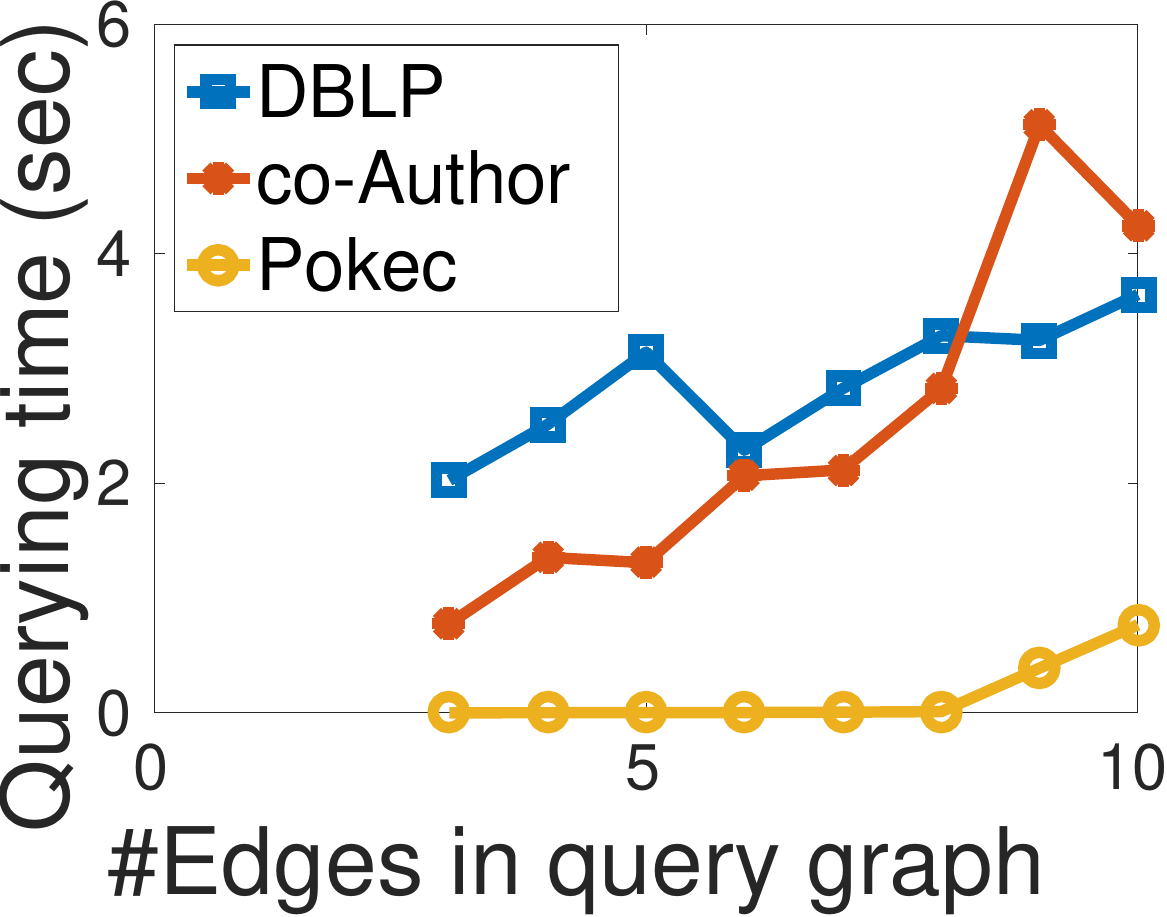}}&
    \subfloat[DBLP \label{fig:TargetSizeEdgeAlternate}]{\includegraphics[width=\imageWidthIcePlot]{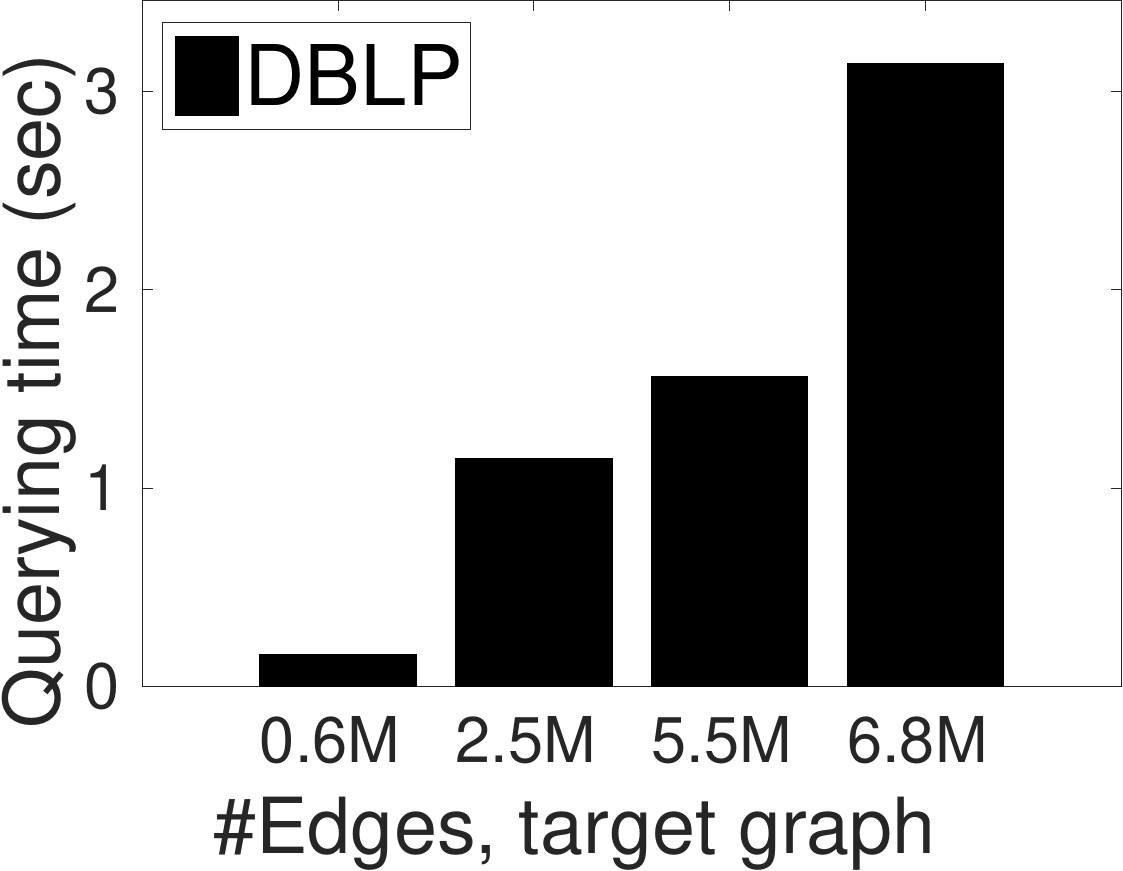}}&
    \subfloat[Pokec \label{fig:TargetSizeEdgePokec}]{\includegraphics[width=\imageWidthIcePlot]{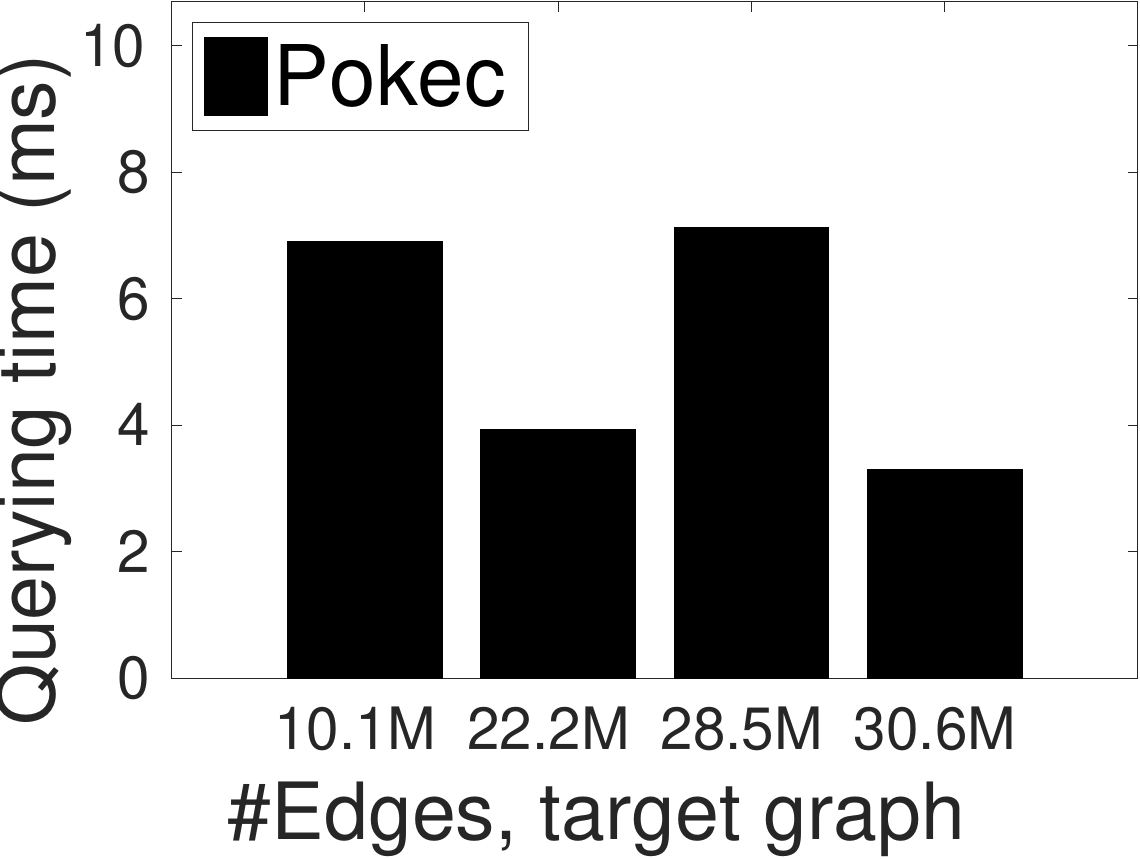}} \\
    \subfloat[DBLP
    \label{fig:TargetSizeAlternate}]{\includegraphics[width=\imageWidthIcePlot]{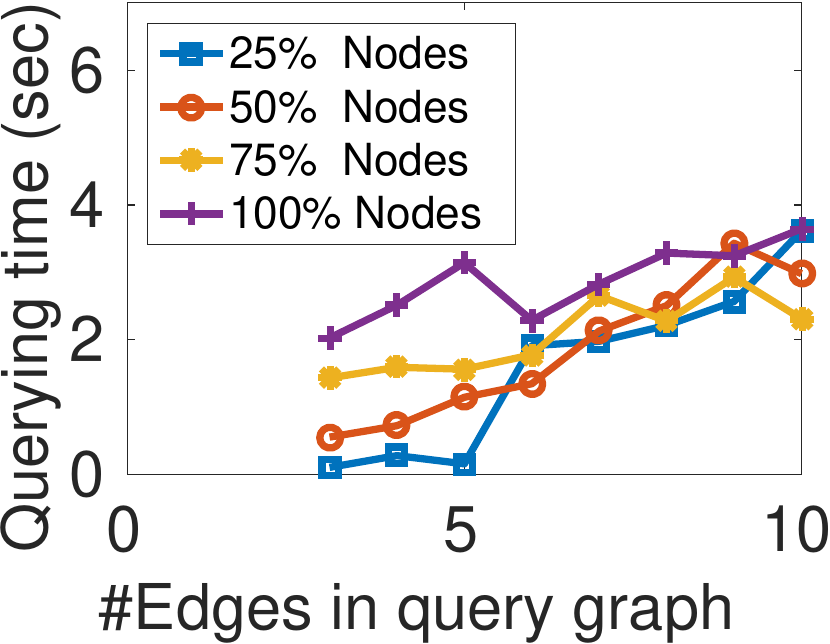}} &
    \subfloat[Pokec \label{fig:TargetSizePokec}]{\includegraphics[width=\imageWidthIcePlot]{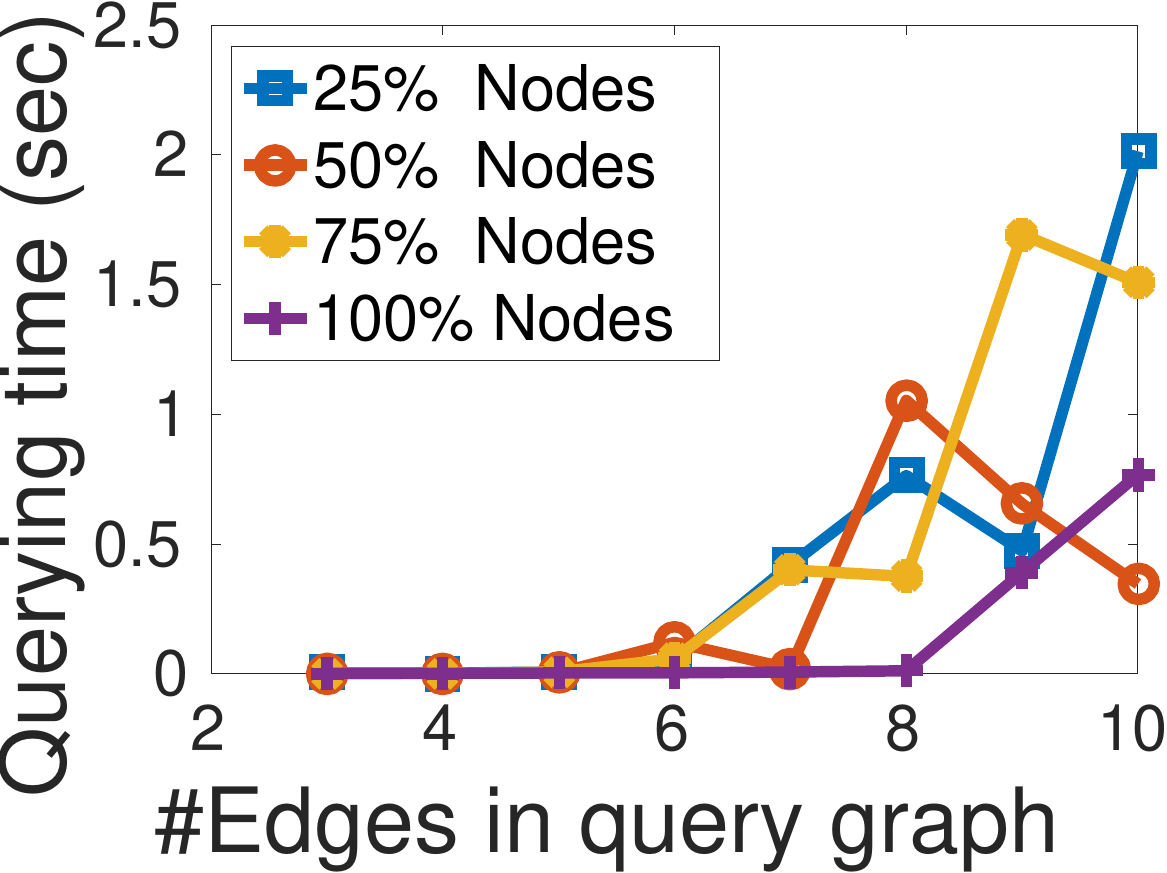}} &
    \subfloat[Large datasets \label{fig:K}]{\includegraphics[width=\imageWidthIcePlot]{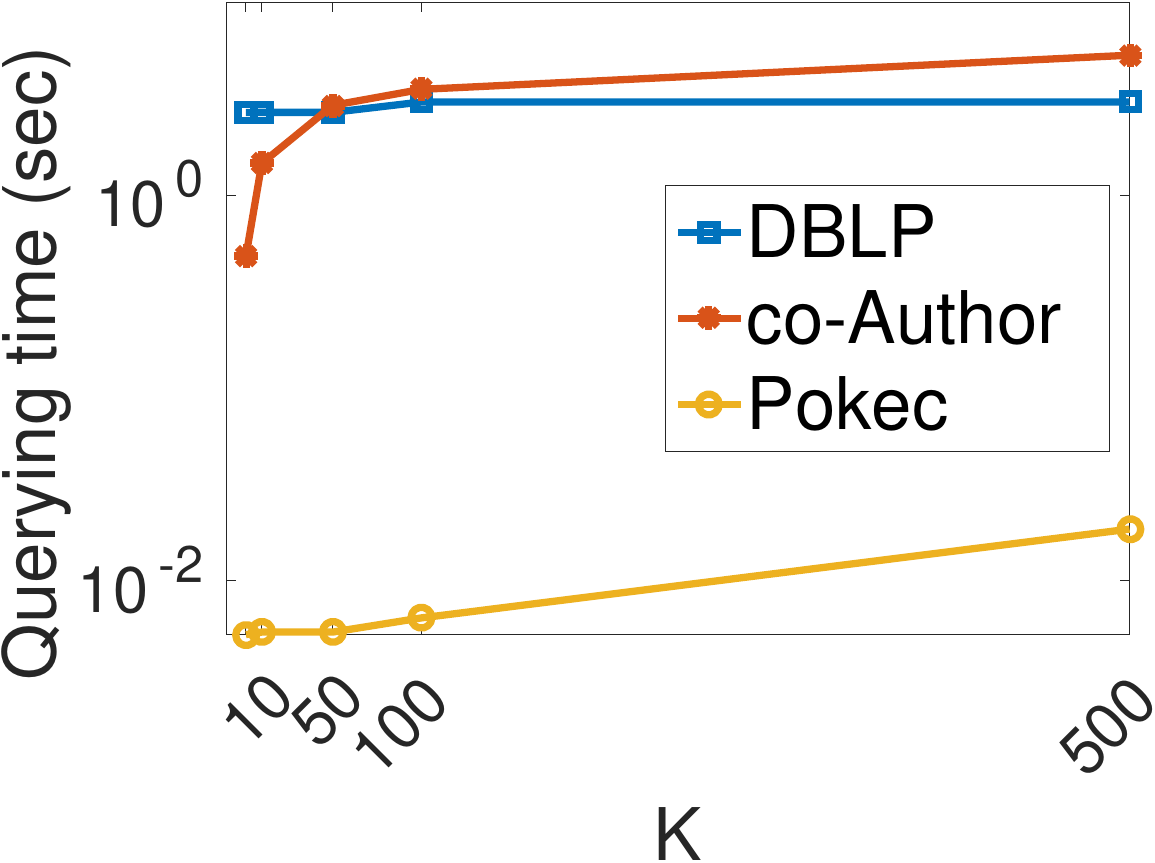}} &
    \subfloat[DBLP \label{fig:IndexConstrctionTime}]{\includegraphics[width=1.2in]{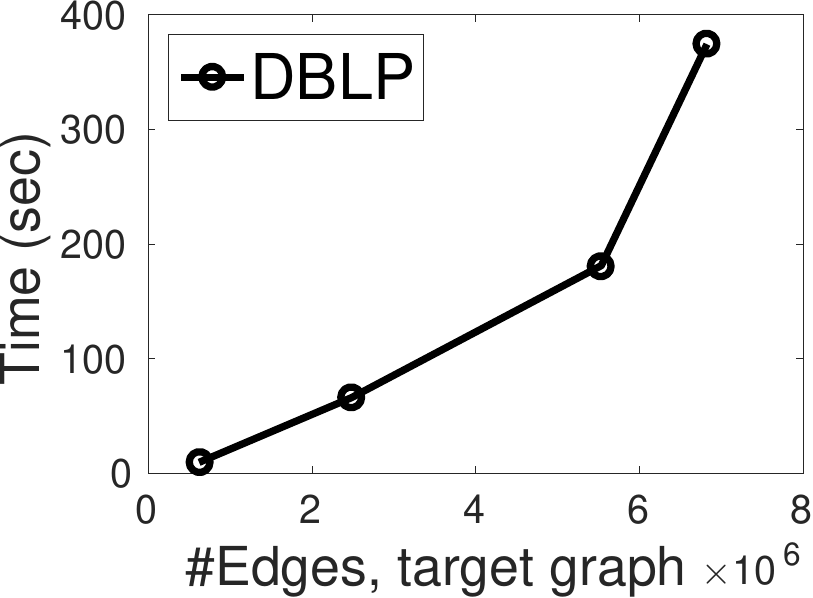}} &
    \subfloat[Pokec \label{fig:IndexConstrctionTimePokec}]{\includegraphics[width=1.2in]{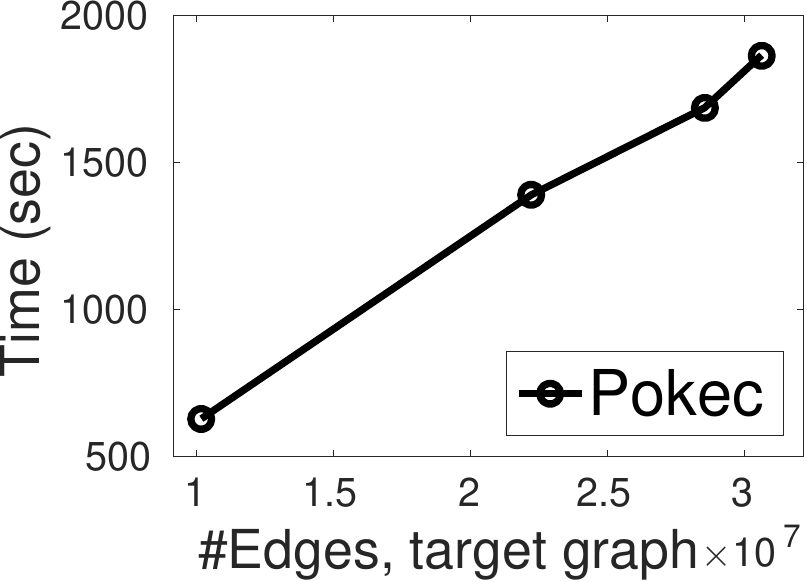}}
  \end{tabular}
  \figcaption{\textbf{(a) Comparison of traditional graph search with the proposed relationship-aware similarity function. (b) Comparison of querying time between RAQ and the baseline technique. (c) Growth of querying time against query graph size and (d-e) target graph size. (f-g) Impact of target graph size on queries of various sizes. (h) Growth of querying time against $k$. (i-j) Growth of index construction time against network size.}}
  \label{(a)}
  \end{figure*}
Interestingly, across all $25$ queries, QBE returned an empty answer set since it did not find any subgraph from IMDb, other than the query itself, which had nodes connected by the same relationships as in the query. This result is not surprising. As we discussed in Sec.~\ref{sec:qbe}, QBE operates in a binary world where two relationships are either identical or different. %This strict requirement is further compounded by the fact that $5$ relationships (corresponding to $5$ node features) exist between each pair of nodes. As a result, the chances of having another subgraph connected by an identical set of relationships is extremely low. 
This result indicates that while QBE is an intuitive and powerful framework for relationship-annotated knowledge graphs, it is not well-suited for graphs where the relationships are encoded through interactions between feature vectors characterizing each node. Owing to this result, the user survey reduced to a comparison among three techniques.

%Moreover, an equally high number of responses ($697 \approx 82\%$) considered the results retrieved using graph edit distance to be the least relevant (rank-3 match) to the query. 
Table~\ref{tab:user_study} summarizes the user responses and in Fig.~\ref{fig:user_study_response_distribution}, we plot the distribution of the received responses within each retrieved rank for each query. An overwhelming $731$ ($86\%$) of the $850$ received user responses deemed RAQ results to be the most relevant (rank-1 match) to the input query. Considering the null hypothesis where any one of the three results is equally likely to be chosen as the preferred one, this observed outcome is highly statistically significant %with a $X^2$ value of $1067.997$ and 
($p$-value of $2\mathrm{e}{-232}$). Moving beyond the analysis of rank-1 responses, we observe a high rank-correlation between the user-responses and the expected outcome, i.e., RAQ, RAQ-Uniform, and GED at ranks 1, 2, and 3 respectively, indicated by the average Spearman's coefficient $\rho=0.803$\cite{spearman}. It is evident from Fig.~\ref{fig:rho_survey2} that $\approx 74\%$ of the user-responses perfectly match the expected output ranking with $\rho=1$.

In addition to analyzing the user-level responses, we performed a query-level analysis as well. Specifically, for each query-rank combination, we identified the algorithm  preferred by the highest number of users. 
%To this end, for each query and retrieved rank we identify the technique that was deemed relevant the highest number of times in the user responses. 
As is clear from Table~\ref{tab:user_study}, RAQ results were considered to be the most relevant (rank-1) for all $25$ ($100\%$) queries.
%, while both RAQ-Uniform and graph edit distance were found to be relevant to ranks 2 and 3 respectively in $23$ ($92\%$) out of $25$ queries.
%We next dive deeper into each query 
%Both the analyses presented above clearly establish the preference of users towards relationship-aware matches. Moreover, it can be concluded that in majority of the cases RAQ results should be ranked at the first position, while RAQ-Uniform and graph edit distance should be ranked at second and third position respectively. To better understand 
%and analyze their impact on the user responses. In Fig.~\ref{fig:user_study_response_distribution}, we plot the distribution of the received responses within each retrieved rank for each query. Fig.~\ref{fig:user_study_response_distribution} shows that for majority of the queries (with $Q12$ and $Q23$ representing the most ideal scenarios) the user-responses prefer RAQ at rank-1, RAQ-Uniform at rank-2, and GED at rank-3. %$Q10$ somewhat deviates from this trend. To better understand the reason behind this behavior, we interviewed a small subset of $10$ users from our user-base, and identified that the users found it hard to disambiguate between the highly similar results produced by RAQ and RAQ-Uniform respectively. %Nevertheless, even then $65\%$ of the user responses favoured RAQ over the baselines. 
%Overall, the user survey resulted in highly encouraging results corroborating the importance of the proposed RAQ paradigm.

\subsection{Case Study: DBLP Dataset}
\label{sec:quality}

In this section, we showcase results from the DBLP co-author dataset and discuss how RAQ is able to identify relationship patterns that node-based distance measures like GED are unable to. Each node in this experiment is characterized by three features, \emph{number of papers published}, \emph{median venue rank}, and \emph{subject area}. The subject area corresponds to the area of the venue where the author has published most of her works. 
%Note that, we do not necessarily present the rank-$1$ matches for the queries. Rather we present the results containing \emph{authors popular in the database community} and thereby being relevant to the usual readers of a database paper. Nevertheless, the top-$5$ matches of each query are provided in Appendix in Figs.~\ref{fig:q1top5}-\ref{fig:q3top5} for the readers' reference. 
%Furthermore, all presented results of the relationship-aware query in this section are within top-10.

The query presented in Fig.~\ref{fig:query1} represents a collaboration pattern comprising prolific authors %-- \emph{Jiawei Han, Philip S. Yu, and Charu C. Aggarwal} -- 
working in the field of \emph{data mining}. Fig.~\ref{fig:result1} presents the matched subgraph by RAQ, which is a collaboration pattern among prolific authors %-- \emph{Magdalena Balazinska, Ugur \c Cetintemel, and Michael Stonebraker} -- 
working in the field of \emph{databases}. As can be seen, all authors in Fig.~\ref{fig:result1} possess the median venue rank as ``A*'' and the subject area as ``DB''. It is evident that RAQ captures the relationship that each group contains prolific authors from the same community. The result presented in Fig.~\ref{fig:result1} is not even within the top-10,000 matches produced by GED. This is natural since the two groups do not contain similar nodes.
%rather the relationship connecting the nodes in each group are similar. We emphasize that we are not using this result to claim RAQ as better than traditional techniques. Rather, the stress is on RAQ's ability to identify useful matches that traditional techniques are unable to.
% under traditional graph matching. Generally, traditional graph matching  quantifies the quality of correspondence between the mapped nodes  across two graphs\cite{ctree,NEMA}. Thus, given an MCS $g$ between a query graph $q$ and a target graph $G$, the traditional similarity between $q$ and $g$ is calculated as the summation of \emph{min-max} similarity (similar to Definition~\ref{def:rv}) between the feature-values of the mapped nodes. For further details on the traditional similarity function please see Appendix~\ref{app:traditional}.

%It is important to note here that a traditional graph similarity (or distance) function would never be able to capture this result. 
%As can be seen, there is negligible similarity between features of the nodes in the two graphs. 
%Node similarity is a basic necessity in all traditional graph matching algorithms. 
%Consequently, regardless of the similarity function used to capture node similarity, these two graphs would always be considered dissimilar. It is therefore not surprising to observe that Fig.~\ref{fig:result1} has a rank beyond $10000$ under the traditional graph matching paradigm

Fig.~\ref{fig:query2} demonstrates RAQ's ability to adapt with the query. More specifically, while the subject area was a preserved relationship in query 1, query 2 presents collaborations among authors with high paper counts, but from diverse backgrounds. The result, Fig~\ref{fig:result2}, is a collaboration pattern among authors from diverse backgrounds as well, but all prolific with a large number of papers. 

%Finally, a third query is shown in Fig.~\ref{fig:query3}. %Similar to the previous two queries, the result, shown in Fig.~\ref{fig:result3.1}, is intuitive and meaningful. The second result of this query, shown in 
%Fig.~\ref{fig:result3.2} unearths a different property of the relationship-aware similarity function: \emph{backward compatibility}. More specifically, this is one of the rare results, where edit distance provides a relatively high rank of $68$. This high rank is not a consequence of traditional matching capturing the relationships, rather a coincidence where the node feature vectors between the query and the result happen to be similar. Now, when the mapped nodes across two graphs are similar, from the formulation of relationship vector, it is guaranteed that the corresponding relationship vectors would also be similar. Consequently, relationship-aware similarity is backward compatible with node-based similarity functions.

\noindent
\textbf{Correspondence between RAQ and edit distance:} In this section, we show that RAQ retrieves results that node-based techniques are unable to find. 
%In the next experiment, we quantify this uniqueness of \emph{RAQ}. 
%This provides ground for an alternate and more meaningful interpretation: \emph{RAQ can identify all the results that are identified as relevant under the traditional similarity metric but not vice versa}. 
We perform a top-$5$ query on random query graphs with sizes varying from $3$ to $10$. For each of the top-$5$ RAQ results, we find its rank using GED. We plot the average rank in GED for each of the query graph sizes (Fig.~\ref{fig:QualitativeQuantitativenh}). Since it is hard to scale beyond top-10,000 in GED, if a RAQ result does not appear within top-10,000, we set its rank to 10,000.
%One of the most important aspect of RAQ, besides its ability to capture context, is backward comparability. RAQ is also able to find subgraphs which are closer in traditional sense. To understand it more, consider the query graph shown in Fig. ~\ref{fig:Query2}. One of the top results corresponding to this is shown in Fig. ~\ref{fig:Result1.1}. The context defined by the query graph requires the area and the rank to be similar, as can be seen in the result we have found a subgraph which also has a similar context. Thing to note about this result is that it could identify a subgraph of authors working in a area different from that of query graph. This fact is reflected in the high value of traditional distance. Another top result is shown in Fig. ~\ref{fig:Result1.2}, this subgraph preserves both the context and also the area of research. As can be seen, both the traditional distance and RAQ distance very low.	
%even for Since in many, $s_t$, from the query graph; post which we perform a range query with a similarity threshold $s_t$. While performing the range query the number of retrieved results were greater than $10,000$ in majority of the cases, therefore, to ensue practicality, we limit the range query result set size to $10,000$. 
The GED rank of the top-5 RAQ matches is close to 10,000 on average. 
 This behavior is not a coincidence. 
Node-based similarity functions match graphs by considering each node as an independent entity and remains blind to the patterns found when the nodes are considered as a group. Consequently, they miss out on similarities that RAQ is able to discover.

%In summary, this section, in conjunction with the user-survey in Sec.~\ref{sec:survey}, highlight the utility of RAQ in surfacing useful results that would otherwise remain hidden. RAQ complements existing graph querying algorithms with a new relationship-aware approach and herein lies the key contribution of our work.

%smeans that none of the top-$5$ results occur in even the top-$10000$ results of traditional search. This phenomenon is also indicated in Figs.~\ref{fig:result1}, ~\ref{fig:result2}, and~\ref{fig:result3.1}.	

\subsection{Efficiency and Scalability}
\label{sec:efficiency}

%After thoroughly establishing the qualitative aspect of RAQ, in this section, we evaluate the efficiency and scalability of RAQ search algorithm. 
We next compare the running time of RAQ with RAQ-uniform and QBE \cite{exemplar1} (Fig.~\ref{fig:runtime}). We omit GED since for query graphs sizes beyond 5 it is impractically slow. RAQ is up to 5 times faster than QBE which suffers due to replication of each edge $d$ times corresponding to $d$ relationships. RAQ is faster than RAQ-uniform, since in RAQ, the relationships weights are skewed. Hence, pruning strategies are more effective than RAQ-uniform, where all relationships are equally important.
 
%First, we quantify the impact of the RAQ search algorithm. Towards that, we perform top-$10$ queries on the Northeast dataset using random query graphs of sizes varying from $3$ to $12$ edges using two procedures. In the first procedure, we use the proposed search algorithm and in the second procedure, we search in bottom-up manner without using any of the index structures and the associated pruning strategies. The result is shown in Fig.~\ref{fig:BaselineDFS_Comparison_NE}. As is evident, the proposed search algorithm imparts a speed-up of up to $1000$ times. 
%The poor performance of baseline technique is because of the complex nature of our problem. The problem by itself is NP-hard, and in addition, since we are looking for maximal common subgraphs, we cannot employ any sophisticated search scheme involving the structure of graph. In addition, the non-importance of labels further increases the search space. 
%Recall, that the problem is NP-complete. RAQ is able to overcome the exponential search space by limiting its search efforts only to those regions in the target graph that contain the best matches to the query with a high likelihood. This is made possible through beam stack search, which provides good seed candidates and then exploiting neighborhood signatures to focus on promising regions of the search space. An alternative view of the same result in terms of speed-up against query graph size is provided in the Appendix in Fig.~\ref{fig:BaselineDFS_SpeedUp_NE}.

Fig.~\ref{fig:QuerySize} demonstrates the growth of querying time against query graph size. As expected, the growth of is exponential since the search space grows exponentially with query graph size. Nonetheless, RAQ
%successfully mitigates the effects of exponential growth through effective pruning mechanisms, which 
limits the running time to a maximum of only 5\,s. 

Notice that RAQ is fastest in Pokec even though Pokec is the largest network containing more than $30$ million edges. The variance in feature values in Pokec is low. Consequently, the chance that a randomly picked subgraph is identical to the query is larger. Owing to this property, the search for the query graph converges quickly. 
%On the other hand, the DBLP datasets are more heterogeneous and therefore provide a tougher challenge. 
Overall, these results establish that RAQ is a fast and practical search mechanism.
%In this three datasets, the baseline technique is unable to scale beyond query graph size of $5$. %The comparison with baseline for smaller query sizes is provided in Fig.~\ref{fig:QuerySize_Baseline_app} in Appendix. Even for smaller query sizes, RAQ is more than three orders of magnitude faster.
%Another observation one can draw from the experiment is the reduction in querying time for query graphs of size 7. This phenomenon will be explained in section ~\ref{experiments:optimization}.
%This fact is more pronounced in Fig. ~\ref{fig:QuerySize_Baseline}, where we also compare the results with our baseline technique. Since the baseline technique did not scale beyond a query size of 5, we have restricted the results for it. 
%Fig. ~\ref{fig:QuerySize} repeats the same experiments on the three larger datasets. We vary the size of the query graph and plot the growth of the querying time in RAQ. Across all three datasets, RAQ finishes in less than 90ms. 
%We study the effects of query graph size, target graph size and query parameter k, on the search time. We also study the memory footprint of the indexing structure.

\begin{figure}[t]
  \begin{tabular}{ccc}
	\subfloat[Pokec \label{fig:IndexSize}]{\includegraphics[width=0.99in]{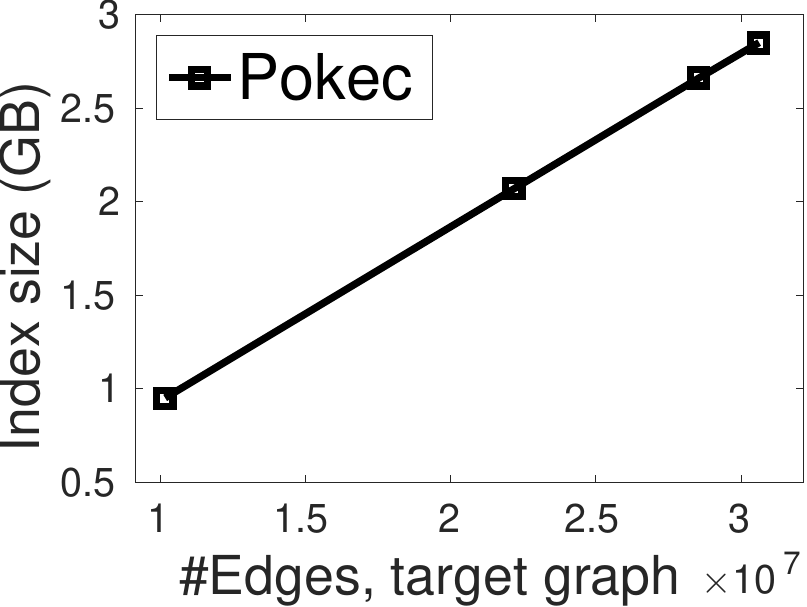}} &
    \subfloat[Beam-width \label{fig:BeamWidth}]{\includegraphics[width=0.99in]{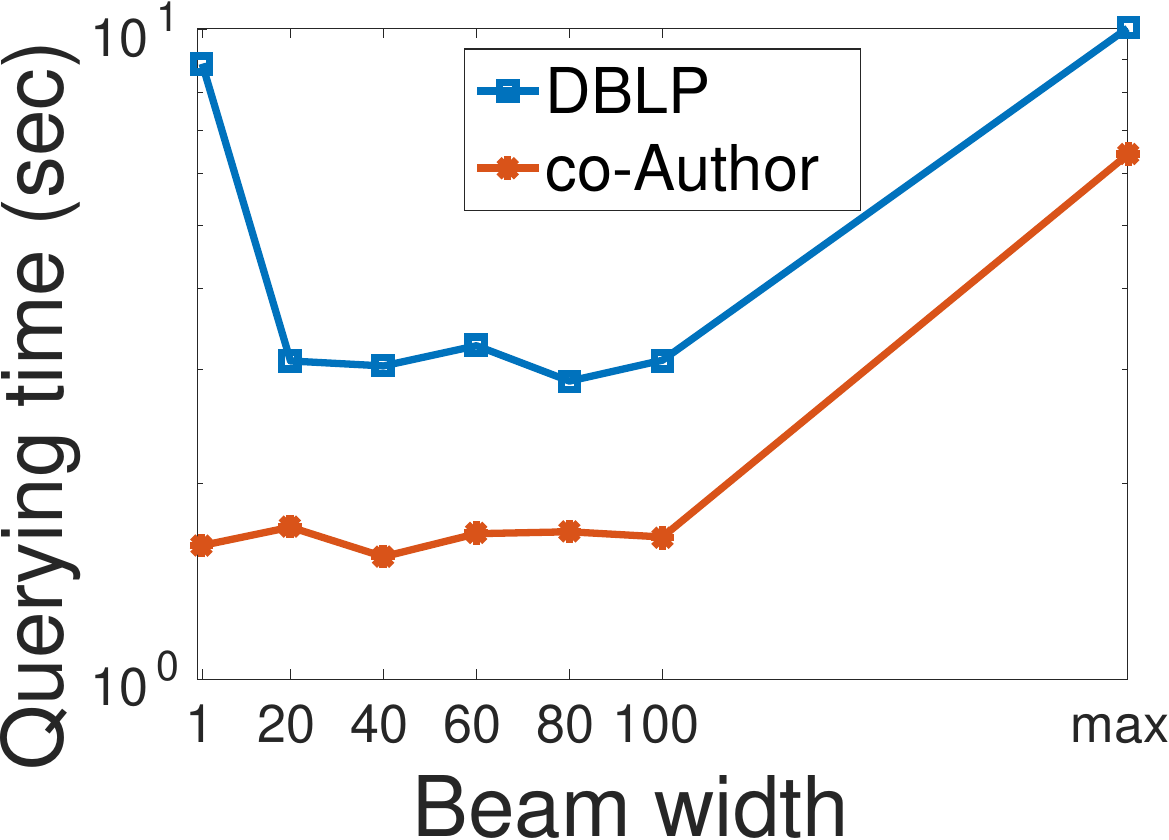}} &
    \subfloat[RWR Signatures \label{fig:Heuristics_Comparison_DBLB_Bar}]{\includegraphics[width=0.99in]{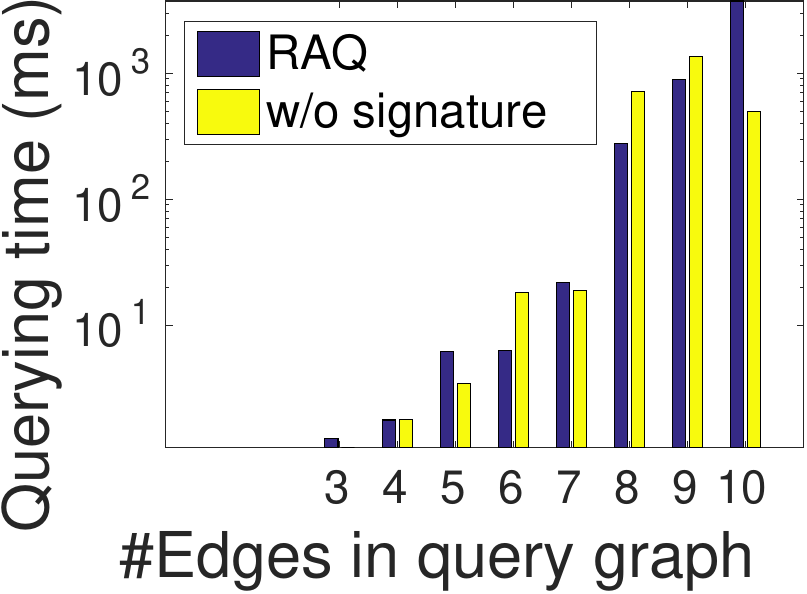}}%&
%    \subfloat[ \label{fig:BaselineBFS_Comparison_NE}]{\includegraphics[width=1.30in]{pdf/BaselineBFS_Comparison_NE_Bar.pdf}} &
  %  \subfloat[\label{fig:BranchingFactor}]{\includegraphics[width=\imageWidthIcePlot]{pdf/BranchingFactor.pdf}} %\\  
  \end{tabular}
\vspace{-0.05in}
  \figcaption{\textbf{(a) Growth of RAQ's memory footprint against network size. Impact of (b) beam width and (c) neighborhood signatures %, and (d) R-tree.% and (e) branching factor of R-tree 
on querying time.
}}
\end{figure}
%The second most important factor influencing the querying time is the target graph size. We study its effect on the two largest datasets: DBLP co-author and Pokec. 
Next, we study the impact of target graph size on the querying time.
For this experiment, we extract four different subgraphs of the entire network covering $25\%$, $50\%$, $75\%$ and $100\%$ of the nodes (and all edges between them) of the entire network (Figs.~\ref{fig:TargetSizeEdgeAlternate},~\ref{fig:TargetSizeEdgePokec})
%To pick a subgraph covering $X\%$ of all nodes in the entire network, we start from an arbitrary node, and perform a random walk till at least $X\%$ of the nodes have been visited at least once. On each of the subgraphs, we plot the growth rate of querying time across various query graph sizes.
%On each of these target graphs we generate random query graphs of sizes varying from 3 to 12 and perform a top-10 query. The results of this can be seen in Fig. ~\ref{fig:TargetSizeAlternate}. 
In DBLP, the querying time increases with increase in target graph size. 
%This is natural since the search space increases exponentially with increase in target graph size. Notice that 
In Pokec, we do not observe a similar trend. Since we employ best-first search, the search procedure stops as soon as the upper bound of an unexplored candidate is smaller than the $k^\text{th}$ most similar subgraph we have already found. While an increase in target graph size increases the search space, it also enhances the chance of the $k^\text{th}$ subgraph being more similar and therefore an earlier termination of the best-first search. Due to this conflicting nature of our search operation, we do not see a consistent trend in the running time against the target graph size. %This could be because the random selection of nodes can generate a subgraph with many similar edges and neighborhoods  and, hence, a larger search space.

Figs.~\ref{fig:TargetSizeAlternate} and ~\ref{fig:TargetSizePokec} study the impact of query graph size on the querying time. In general, we see any upward trend in the querying time as the size of the query graph increases.
%Here, we observe a more linear growth rate similar to the results of Fig.~\ref{fig:QuerySize}.
%because target graph size has a direct correlation to the size of search space. We observe RAQ framework has not introduced any overhead of its own and has  However, due to the effective pruning mechanisms of RAQ, we observe a more linear growth in querying time than exponential. 
%To emphasize this, we have performed top-7 queries on DBLP co-author networks of different size. The results of this can be seen in Fig. ~\ref{fig:TargetSizeEdgeAlternate}.
%The final factor influencing querying time is $k$. 

Fig.~\ref{fig:K} presents the variation of querying time across various values of $k$.
%As expected, the querying time increases with increase in $k$. %What is more interesting to observe however is 
We notice that the querying time flattens out very rapidly. This is due to the fact that the number of subgraphs explored till the answer set converges remains relatively same even for larger values of $k$.

\noindent
\textbf{Indexing Costs: }%After a thorough investigation of the querying time, 
%We next shift our focus to offline indexing costs. 
%As a first step, we plot the time taken to construct the RAQ index structure. The index structure contains two components: R-tree on relationship vectors and the neighborhood signatures of all edges. 
Figs.~\ref{fig:IndexConstrctionTime},~\ref{fig:IndexConstrctionTimePokec} present the results of indexing costs on DBLP and Pokec. Even on Pokec, which contains more than $30$ million edges, RAQ completes index construction within $33$ minutes. %Note that index construction is a one-time offline activity and thus not a critical aspect. 
%The memory footprint of the index structure is a relatively more important aspect. 
RAQ index structures have a linear space complexity as reflected in Fig.~\ref{fig:IndexSize} where we depict the memory footprint.
%To extract a subgraph of desired size from Pokec, we use the same random walk based method employed earlier in Figs~\ref{fig:TargetSizeAlternate}-\ref{fig:TargetSizePokec}. 
%The linear growth rate of memory footprint is a highly desirable characteristic, which allows RAQ to be applied on million-scaled datasets. As can be seen, even for a graph having $30$ million edges, we consume less than $3GB$ of memory. Overall, these results establish that RAQ is not only scalable in querying time, but also in its index construction costs.

\subsection{Optimization}
\label{experiments:optimization}

The efficiency of the search algorithm RAQ is dependent on two major components: 
%\emph{R-tree} for similar edge search, 
\emph{neighborhood signatures} for identifying similar neighborhoods and \emph{beam-stack search} to grow and prune initial candidates. In this section, we analyze their impact on RAQ. %In addition, we look into the effects of the internal parameters used by RAQ.

Beam-width controls the number of simultaneous subgraphs we grow to form isomorphic matches to the query. We vary this parameter and plot the corresponding querying times in %A beam-width of $1$ would make the search work like a depth-first search. On the other hand, an infinite beam width would degenerate to best-first search. To study the effect of this parameter, we run top-$10$ query with beam-width varying from 1 to $\infty$. 
%The random query graphs are all of size 7. 
Fig.~\ref{fig:BeamWidth}.
A beam width of $1$
reduces to depth-first search,
where an initial bad choice needs to be explored till the end before
moving to a different one.
A larger beam width allows other
promising candidates explored earlier.
A very large value, however, keeps switching too often among the candidates.
Overall, a beam width between $40$ and $60$
provides the highest speed-up.

We study the importance of neighborhood signatures. We compare the time taken for top-$k$ search using two techniques: RAQ with neighborhood prioritization through signatures and RAQ without neighborhood prioritization. 
%We generate random query graphs of sizes varying from 3 to 10 and perform top-10 query. 
Fig.~\ref{fig:Heuristics_Comparison_DBLB_Bar} presents the results in the Pokec
dataset. We observe that for smaller query graphs, RAQ without neighborhood
optimization performs better. However, for larger query graphs neighborhood, typically,
 optimization helps. For smaller query graphs, the
neighborhood of a query edge itself is very small and, thus, they lack much
discriminative information.
%Thus, the extra computation time taken to construct the signatures on the query graph does not provide any additional pruning capability. 
When the query size grows, the neighborhood of an edge is larger and the chances of getting stuck in a local maxima increases. Signatures allows us to escape these regions. %and consequently we observe the trend visible in Fig.~\ref{fig:Heuristics_Comparison_DBLB_Bar}.

\section{Conclusions}

In this paper, we addressed the problem of graph querying under a novel
\emph{Relationship-Aware Querying (RAQ)}
paradigm. RAQ captures the relationships present among nodes in the query graph
and uses these relationships as an input to the similarity function. Majority
of the existing techniques consider each node as an independent entity and,
hence, remain blind to the patterns that exist among them when considered as a
whole. RAQ not only mines these patterns in the form of relationships but also
weighs their importance in proportion to their statistical significance.
An overwhelming majority of $86\%$
of the users surveyed preferred RAQ over existing metrics. To address the computational
challenges posed by graph querying, we designed a flexible searching algorithm,
which performs beam-stack search on an R-tree to prune the search space.
Empirical studies on real-world graph datasets established that RAQ is scalable
and fast. Overall, RAQ opens a new
door in the graph querying literature by surfacing useful results that would
otherwise remain hidden.
%thereby providing a speed-up of up to \emph{three orders} of magnitude over a baseline strategy. 
%In future, we would like to come up with a distributed version of our algorithms and explore interesting applications like knowledge graph search.

\pagebreak

\bibliographystyle{abbrv}
\balance
\bibliography{raq}

\begin{thebibliography}{10}

\bibitem{core_ranking}
Core rankings portal.
\newblock \url{http://portal.core.edu.au/conf-ranks/}.

\bibitem{dblp}
{DBLP: Computer Science Bibliography}.
\newblock \url{http://dblp.uni-trier.de/}.

\bibitem{facebook}
{Facebook Graph Search}.
\newblock \url{https://en.wikipedia.org/wiki/Facebook_Graph_Search}.

\bibitem{google}
{Google Knowledge Graph}.
\newblock
  \url{https://www.google.com/intl/es419/insidesearch/features/search/knowledge.html}.

\bibitem{imdb}
{IMDB: The Internet Movie Data Base}.
\newblock \url{https://www.imdb.com/interfaces/}.

\bibitem{snap}
{SNAP Datasets}.
\newblock \url{https://snap.stanford.edu/data/}.

\bibitem{graphchi}
A.~Arora, M.~Sachan, and A.~Bhattacharya.
\newblock Mining {S}tatistically {S}ignificant {C}onnected {S}ubgraphs in
  {V}ertex {L}abeled {G}raphs.
\newblock In {\em SIGMOD}, pages 1003--1014, 2014.

\bibitem{dbpedia}
S.~Auer, C.~Bizer, G.~Kobilarov, J.~Lehmann, R.~Cyganiak, and Z.~Ives.
\newblock Dbpedia: A nucleus for a web of open data.
\newblock In {\em ISWC}, pages 722--735, 2007.

\bibitem{indexingbook}
A.~Bhattacharya.
\newblock {\em Fundamentals of Database Indexing and Searching}.
\newblock CRC Press, 2014.

\bibitem{binning}
M.~Biba, F.~Esposito, S.~Ferilli, N.~Di~Mauro, and T.~M.~A. Basile.
\newblock Unsupervised discretization using kernel density estimation.
\newblock In {\em IJCAI}, pages 696--701, 2007.

\bibitem{freebase}
K.~Bollacker, C.~Evans, P.~Paritosh, T.~Sturge, and J.~Taylor.
\newblock Freebase: A collaboratively created graph database for structuring
  human knowledge.
\newblock In {\em SIGMOD}, pages 1247--1250, 2008.

\bibitem{mcg_similarity1}
H.~Bunke and K.~Shearer.
\newblock A graph distance metric based on the maximal common subgraph.
\newblock {\em Pattern recognition letters}, 19(3):255--259, 1998.

\bibitem{mywww}
V.~Chaoji, S.~Ranu, R.~Rastogi, and R.~Bhatt.
\newblock Recommendations to boost content spread in social networks.
\newblock In {\em WWW}, pages 529--538, 2012.

\bibitem{graphquerying_chi}
S.~Dutta, P.~Nayek, and A.~Bhattacharya.
\newblock Neighbor-aware search for approximate labeled graph matching using
  the chi-square statistics.
\newblock In {\em WWW}, pages 1281--1290, 2017.

\bibitem{deltacon}
C.~Faloutsos, D.~Koutra, and J.~T. Vogelstein.
\newblock {DELTACON:} {A} principled massive-graph similarity function.
\newblock In {\em SDM}, pages 162--170, 2013.

\bibitem{mcg_similarity2}
M.-L. Fern{\'a}ndez and G.~Valiente.
\newblock A graph distance metric combining maximum common subgraph and minimum
  common supergraph.
\newblock {\em Pattern Recognition Letters}, 22(6):753--758, 2001.

\bibitem{rtree}
A.~Guttman.
\newblock R-trees: A dynamic index structure for spatial searching.
\newblock In {\em SIGMOD}, pages 47--57, 1984.

\bibitem{ctree}
H.~He and A.~K. Singh.
\newblock Closure-tree: An index structure for graph queries.
\newblock In {\em ICDE}, 2006.

\bibitem{exemplar2}
N.~Jayaram, A.~Khan, C.~Li, X.~Yan, and R.~Elmasri.
\newblock Querying knowledge graphs by example entity tuples.
\newblock {\em TKDE}, 27(10):2797--2811, Oct 2015.

\bibitem{stattests}
G.~K. Kanji.
\newblock {\em 100 Statistical Tests}.
\newblock Sage, 2006.

\bibitem{NESS}
A.~Khan, N.~Li, X.~Yan, Z.~Guan, S.~Chakraborty, and S.~Tao.
\newblock Neighborhood based fast graph search in large networks.
\newblock In {\em SIGMOD}, pages 901--912, 2011.

\bibitem{NEMA}
A.~Khan, Y.~Wu, C.~C. Aggarwal, and X.~Yan.
\newblock Nema: Fast graph search with label similarity.
\newblock {\em PVLDB}, 6(3):181--192, 2013.

\bibitem{survey}
S.~Kotsiantis and D.~Kanellopoulos.
\newblock Discretization techniques: A recent survey.

\bibitem{exemplar3}
M.~Lissandrini, D.~Mottin, T.~Palpanas, and Y.~Velegrakis.
\newblock Multi-example search in rich information graphs.
\newblock In {\em ICDE}, 2018.

\bibitem{exemplar}
D.~Mottin, M.~Lissandrini, Y.~Velegrakis, and T.~Palpanas.
\newblock Exemplar queries: Give me an example of what you need.
\newblock {\em PVLDB}, 7(5):365--376, 2014.

\bibitem{exemplar1}
D.~Mottin, M.~Lissandrini, Y.~Velegrakis, and T.~Palpanas.
\newblock Exemplar queries: a new way of searching.
\newblock {\em VLDBJ}, 25(6):741--765, Dec 2016.

\bibitem{resling}
D.~Natarajan and S.~Ranu.
\newblock A scalable and generic framework to mine top-k representative
  subgraph patterns.
\newblock In {\em 2016 IEEE 16th International Conference on Data Mining
  (ICDM)}, pages 370--379. IEEE, 2016.

\bibitem{reslingj}
D.~Natarajan and S.~Ranu.
\newblock Resling: a scalable and generic framework to mine top-k
  representative subgraph patterns.
\newblock {\em Knowledge and Information Systems}, 54(1):123--149, 2018.

\bibitem{node_embedding_similarity}
G.~Nikolentzos, P.~Meladianos, and M.~Vazirgiannis.
\newblock Matching node embeddings for graph similarity.
\newblock In {\em AAAI}, pages 2429--2435, 2017.

\bibitem{pearson}
K.~Pearson.
\newblock Note on regression and inheritance in the case of two parents.
\newblock {\em Proceedings of the Royal Society of London}, 58:240--242, 1895.

\bibitem{pgraphsig}
S.~Ranu, B.~T. Calhoun, A.~K. Singh, and S.~J. Swamidass.
\newblock Probabilistic substructure mining from small-molecule screens.
\newblock {\em Molecular Informatics}, 30(9):809--815, 2011.

\bibitem{nbindex}
S.~Ranu, M.~Hoang, and A.~Singh.
\newblock Answering top-k representative queries on graph databases.
\newblock In {\em Proceedings of the 2014 ACM SIGMOD international conference
  on Management of data}, pages 1163--1174. ACM, 2014.

\bibitem{graphsig}
S.~Ranu and A.~K. Singh.
\newblock Graphsig: A scalable approach to mining significant subgraphs in
  large graph databases.
\newblock In {\em 2009 IEEE 25th International Conference on Data Engineering},
  pages 844--855. IEEE, 2009.

\bibitem{graphsig_jcim}
S.~Ranu and A.~K. Singh.
\newblock Mining statistically significant molecular substructures for
  efficient molecular classification.
\newblock {\em Journal of chemical information and modeling},
  49(11):2537--2550, 2009.

\bibitem{edbt}
S.~Ranu and A.~K. Singh.
\newblock Indexing and mining topological patterns for drug discovery.
\newblock In {\em Proceedings of the 15th International Conference on Extending
  Database Technology}, pages 562--565. ACM, 2012.

\bibitem{chi_square}
T.~R.~C. Read and N.~A.~C. Cressie.
\newblock {\em Goodness-of-fit Statistics for Discrete Multivariate Data}.
\newblock Springer, 1988.

\bibitem{stringchi}
M.~Sachan and A.~Bhattacharya.
\newblock Mining statistically significant substrings using the chi-square
  statistic.
\newblock In {\em PVLDB}, pages 1052--1063, 2012.

\bibitem{isorank}
R.~Singh, J.~Xu, and B.~Berger.
\newblock Global alignment of multiple protein interaction networks with
  application to functional orthology detection.
\newblock In {\em PNAS}, pages 12763--12768, 2008.

\bibitem{spearman}
C.~Spearman.
\newblock The proof and measurement of association between two things.
\newblock {\em The American journal of psychology}, 15(1):72--101, 1904.

\bibitem{saga}
Y.~Tian, R.~C. Mceachin, C.~Santos, D.~J. States, and J.~M. Patel.
\newblock Saga: A subgraph matching tool for biological graphs.
\newblock {\em Bioinf.}, 23(2):232--239, 2007.

\bibitem{tale}
Y.~Tian and J.~M. Patel.
\newblock Tale: A tool for approximate large graph matching.
\newblock In {\em ICDE}, pages 963--972, 2008.

\bibitem{best_effort}
H.~Tong, C.~Faloutsos, B.~Gallagher, and T.~Eliassi-Rad.
\newblock Fast best-effort pattern matching in large attributed graphs.
\newblock In {\em KDD}, pages 737--746, 2007.

\bibitem{rwr}
H.~Tong, C.~Faloutsos, and J.-Y. Pan.
\newblock Fast random walk with restart and its applications.
\newblock In {\em ICDM}, pages 613--622, 2006.

\bibitem{acl}
N.~Voskarides, E.~Meij, M.~Tsagkias, M.~de~Rijke, and W.~Weerkamp.
\newblock Learning to explain entity relationships in knowledge graphs.
\newblock In {\em ACL}, 2015.

\bibitem{leap}
X.~Yan, H.~Cheng, J.~Han, and P.~S. Yu.
\newblock Mining significant graph patterns by leap search.
\newblock In {\em Proceedings of the 2008 ACM SIGMOD International Conference
  on Management of Data}, pages 433--444, 2008.

\bibitem{rwr1}
W.~Yu and X.~Lin.
\newblock Irwr: incremental random walk with restart.
\newblock In {\em SIGIR}, pages 1017--1020, 2013.

\bibitem{ged_similarity2}
Z.~Zeng, A.~K.~H. Tung, J.~Wang, J.~Feng, and L.~Zhou.
\newblock Comparing stars: On approximating graph edit distance.
\newblock In {\em PVLDB}, pages 25--36, 2009.

\bibitem{ged_similarity1}
W.~Zheng, L.~Zou, X.~Lian, D.~Wang, and D.~Zhao.
\newblock Graph similarity search with edit distance constraint in large graph
  databases.
\newblock In {\em CIKM}, pages 1595--1600, 2013.

\bibitem{beamStackSearch}
R.~Zhou and E.~A. Hansen.
\newblock Beam-stack search: Integrating backtracking with beam search.
\newblock In {\em ICAPS}, pages 90--98, 2005.

\end{thebibliography}

\pagebreak

\end{document}